\documentclass[a4paper,11pt]{article} 
\pdfoutput=1

\usepackage{jheppub}

\usepackage[utf8]{inputenc}
 
\usepackage{slashed}

\usepackage{placeins}

\newcommand{\abs}[1]{\left |#1 \right|}
\newcommand{\norm}[1]{\lVert #1 \rVert} 

\newcommand{\Str}{\operatorname{Str}}
\newcommand{\tr}{\operatorname{tr}}
\newcommand{\diag}{\operatorname{diag}}
\newcommand{\Diag}{\operatorname{Diag}}
\newcommand{\adj}{\operatorname{adj}}
\newcommand{\res}{\operatorname{res}}

\renewcommand{\vec}{\mathbf}
\newcommand{\mat}{\boldsymbol}
\newcommand{\ten}{\mat}

\usepackage{xcolor}
 	\definecolor{cardinal}{rgb}{0.77, 0.12, 0.23}

\title{\boldmath Minima of classically scale-invariant potentials}

\author[1]{Kristjan Kannike,\note{Corresponding author.}}
\author{Kaius Loos,}
\author{Luca Marzola}

\affiliation{National Institute of Chemical Physics and Biophysics, \\ R\"{a}vala 10, 10143 Tallinn, Estonia}

\emailAdd{Kristjan.Kannike@cern.ch} 
\emailAdd{kaius.loos@gmail.com}
\emailAdd{Luca.Marzola@cern.ch}

\abstract{We propose a new formalism to analyse the extremum structure of scale-invariant effective potentials. The problem is stated in a compact matrix form, used to derive general expressions for the stationary point equation and the mass matrix of a multi-field RG-improved effective potential. Our method improves on (but is not limited to) the Gildener-Weinberg approximation and identifies a set of conditions that signal the presence of a radiative minimum. When the conditions are satisfied at different scales, or in different subspaces of the field space, the effective potential has more than one radiative minimum.  We illustrate the method through simple examples and study in detail a Standard-Model-like scenario where the potential admits two radiative minima. Whereas we mostly concentrate on biquadratic potentials, our results carry over to the general case by using tensor algebra.}

 \arxivnumber{2011.12304}

\begin{document} 
\maketitle
\flushbottom

\section{Introduction}
\label{sec:intro}

The emergence of radiative minima in scale-invariant potentials is relevant to many problems in contemporary physics. The underlying principle of classical scale invariance, which forbids hard-coded length or mass scales to appear in the action of a fundamental theory of Nature, has been generally adopted  to motivate the smallness of a parameter in the theory. For instance, scale invariance has been used to solve the hierarchy problem posed by the observed Higgs boson mass and to counter the vacuum meta-stability of the Standard Model \cite{Bardeen:1995kv,Heikinheimo:2013fta,Gabrielli:2013hma,Englert:2013gz,Kannike:2016wuy,AlexanderNunneley:2010nw,Foot:2007as,Foot:2007iy,Foot:2010av,Foot:2011et,Antipin:2013exa,Endo:2015ifa,Endo:2016koi,Helmboldt:2016mpi,Lewandowski:2017wov,Lee:2012jn}. Similarly, the framework has also been invoked within neutrino phenomenology \cite{Foot:2007ay,Meissner:2006zh,Kang:2014cia,Lewandowski:2017wov,Iso:2009ss,Khoze:2013oga,Ahriche:2015loa} and within dark energy model building \cite{Shaposhnikov:2008xb,Foot:2010et}. 
Another topic that certainly draws from this fascinating principle is the cosmological inflation \cite{Albrecht:1982wi,Ellis:1982dg,Ellis:1982ws,Linde:1981mu,Linde:1982zj,Khoze:2013uia,Kannike:2014mia,Kannike:2015apa,Kannike:2015kda,Karam:2015jta,Barrie:2016rnv,Tambalo:2016eqr,Marzola:2016xgb,Karam:2017rpw,Racioppi:2017spw,Karam:2018mft,Gialamas:2020snr}, since the flat potentials used to describe the  expansion epoch easily allow quantum corrections to determine the properties of the theory. The possible presence of radiatively generated minima in scalar potentials has also been investigated in other branches of cosmology, for example in connection with phase transitions and gravitational wave production  \cite{Espinosa:2008kw,Randall:2006py,Marzola:2017jzl,Sannino:2015wka,Huang:2020bbe}, or within models of dark matter \cite{Foot:2010av,Ishiwata:2011aa,Steele:2013fka,Altmannshofer:2014vra,Benic:2014aga,Kang:2014cia,Guo:2015lxa,Ghorbani:2015xvz,Wang:2015cda,Karam:2016rsz,Helmboldt:2016mpi,Lewandowski:2017wov}. 

The essence of scale invariance is that the interactions between fields are also responsible for the emergence of all of the observed scales \cite{Coleman:1973jx,Gildener:1976ih}. Remarkably, this is a natural achievement of quantum theories, where the fluctuations need \textit{not} respect scale symmetry even when the action contains no explicit scale.

The appearance of divergences in loop diagrams, for instance, forces the introduction of an arbitrary mass scale in the problem, required for the regularisation of the problematic contributions. Because observables cannot depend on an arbitrary parameter, the explicit dependence of a process on this quantity must be compensated by an implicit dependence of the involved physical parameters. This is the essence of the Callan-Symanzik equation, which determines the evolution of couplings, masses and fields with the renormalisation scale, resulting in the loss of scale symmetry at the quantum level. That quantum fluctuations generally violate scale invariance is also demonstrated by the mechanism of dimensional transmutation \cite{Coleman:1973jx}. In this case, the quantum corrections induce the spontaneous symmetry breaking of the theory by producing a radiatively generated minimum in an otherwise trivial scalar potential. Because the minimisation equation relates the resulting vacuum expectation value (VEV) of the scalar field to the quartic coupling of its potential, the mechanism effectively allows to trade one parameter for the other.    

In this paper, we concern ourselves with the impact of quantum corrections on the extremum structure of scale-invariant potentials, proposing a new mathematical framework to gauge their contribution. The central object of our analysis is the effective potential of a generic scale-invariant theory, whose  minima determine the vacuum state and the spontaneous breaking of symmetries. The computation of the minima of the effective potential is generally a cumbersome task: On top of the dimensionality of the problem that scales with the number of scalar fields, the analysis is complicated by the nature of the radiative corrections. In the absence of scale invariance, such contributions seldom modify the properties of the theory determined by the dominant tree-level term. Within scale-invariant models, instead, these corrections often constitute the leading order contribution. In multi-scalar models, the analysis of the effective potential is also hindered by potentially large quantum corrections which depend on different mass scales. These enter the problem through logarithms of the ratios of mass terms to a common renormalisation scale. Consequently, minimising all quantum corrections in the region of interest by choosing a suitable value for this parameter is a straightforward matter only when the effective potential exhibits a single scale. More involved scenarios must then rely on different approaches to ensure that perturbativity is retained \cite{EINHORN1984261,FORD199317,Bando:1992wy,Ford:1994dt,Ford:1996yc,Ford:1996hd,Steele:2014dsa,Casas:1998cf,Manohar:2020nzp}. Scale-invariant theories are not immune to this problem, as several mass scales could result from non-degenerate field-dependent scalar, vector or fermion masses. To cope with the difficulties posed by several different scales, we use the renormalisation group (RG) improvement method proposed in~\cite{Chataignier:2018aud,Chataignier:2018kay}, but our formalism is not limited to this.

In what follows, we thus present a new approach to the problem of radiative minima of a classically scale-invariant potential based on the method of \cite{Kannike:2019upf} (successfully used in \cite{Dias:2020ryz}). We put particular emphasis on biquadratic potentials because they comprise the lion's share of models in the literature. That the Standard Model Higgs boson transforms as an $SU(2)_L$ doublet quite naturally leads to biquadratic potentials when considering portal couplings to new scalar fields that transform non-trivially under gauge transformations. Other possibilities encompass $\mathbb{Z}_{2}$ mirror symmetries, $O(n)$-symmetric and biconical theories. Our strategy is to write a biquadratic potential in terms of the matrix of quartic couplings and the scalar field vector, using a simple formalism that adds to understanding by building on the geometric intuition and is easily implemented with any modern computer algebra system. As a first result, we obtain general expressions for the stationary point equation, its solution and the mass matrix in a compact way that does not depend on the number of scalar fields. The minimum equation is solved via a systematic iteration procedure, whose lowest order yields the Gildener-Weinberg approximation. Not limited to biquadratic potentials, most of our conclusions also apply to generic potentials via tensor algebra. 

Our second main result concerns scenarios where the classically scale-invariant effective potential admits multiple radiatively generated minima.  Although the possibility was originally mentioned in Ref.~\cite{Gildener:1976ih}, so far it has been practically neglected.
 In particular, we determine conditions that the coupling matrix has to satisfy in a minimum and identify semi-analytical criteria that assess the presence of additional minima in the effective potential. We explicitly demonstrate the possibility by using a toy model that incorporates two complex scalar fields, two Weyl fermions and an $SU(2)$ gauge symmetry. Because our toy model is essentially a bare-bones version of the Standard Model extended by a new scalar and the scalar potentials in most such models are biquadratic, we expect our results to be relevant for future studies of cosmological phase transitions and gravitational wave signals in scale-invariant models.

The paper is organised as follows. We start by surveying general and biquadratic tree-level scalar potentials and their properties in section~\ref{sec:V:tree}. The effective potential and its renormalisation group improvement is reviewed in section~\ref{sec:V:eff}. In section~\ref{sec:V:eff:min} we discuss the conditions for the effective potential to develop a minimum.  Analytical and numerical examples of minimum solutions are presented in section~\ref{sec:examples}, while in section~\ref{sec:multi:min} we present detailed scenarios and discuss the criteria for an effective potential to have multiple non-trivial minima. We conclude in section~\ref{sec:concl}.

\section{Tree-level potential}
\label{sec:V:tree}

The generic renormalisable quartic potential of $n$ real scalar fields $\phi_{i}$, collected in the vector $\vec{\Phi}$, is given by
\begin{equation}
  V^{(0)} = \frac{1}{4!} \sum_{_{i,j,k,l}} \lambda_{ijkl} \phi_{i} \phi_{j} \phi_{k} \phi_{l},
\label{eq:pot:gen}
\end{equation}
where $\lambda_{ijkl}$ is the symmetrised tensor of quartic couplings.\footnote{For gauge multiplets it can be advantageous to write the potential in terms of gauge invariants such as norms of fields (see e.g. \cite{Bednyakov:2018cmx} using the formalism of \cite{Ivanov:2005hg,Maniatis:2006fs}).}

For biquadratic potentials, the expression simplifies to
\begin{equation}
  V^{(0)} = \sum_{i,j} \phi_i^2 \lambda_{ij} \phi_j^2 
  = (\vec{\Phi}^{\circ 2})^T \! \mat{\Lambda} \vec{\Phi}^{\circ 2},
\label{eq:pot:biq}
\end{equation}
where $\mat{\Lambda}$ is the symmetric quartic coupling matrix. The $\circ$ symbol denotes the \emph{Hadamard product}, defined as the element-wise product of matrices of same dimensions: $(\mat{A} \circ \mat{B})_{ij} = A_{ij} B_{ij}$. The \emph{Hadamard power} of a matrix $\mat{A}$ is then $(\mat{A}^{\circ n})_{ij} = A_{ij}^{n}$. For example, the elements of the Hadamard square of the field vector, $\vec{\Phi}^{\circ 2}$, are given by $\phi_{i}^{2}$. Biquadratic potentials~\eqref{eq:pot:biq} admit a $\mathbb{Z}_{2}^{n}$ group, so each scalar $\phi_{i}$ is odd under a private $(\mathbb{Z}_{2})_{i}$ symmetry.

The radial coordinate $\varphi$ in the field space is given by the norm of $\vec{\Phi}$, which can be expressed as
\begin{equation}
  \varphi^{2} \equiv \vec{\Phi}^{T} \vec{\Phi} = \vec{e}^{T} \vec{\Phi}^{\circ 2},
\label{eq:Phi:norm}  
\end{equation}
where $\vec{e}= (1, \ldots, 1)^{T}$, the vector of ones, is an identity element of the Hadamard product. Note that the relation~\eqref{eq:Phi:norm} is quadratic in the field $\vec{\Phi}$ but only linear in its square $\vec{\Phi}^{\circ 2}$. We can use the vector $\vec{e}$ to construct identity elements of the Hadamard products of higher dimension, for instance the matrix $\vec{e} \vec{e}^{T}$ with elements all equal to unity.

The tree-level field-dependent scalar mass matrix for the biquadratic potential~\eqref{eq:pot:biq} is given by
\begin{equation}
\begin{split}
  \mat{m}_S^2 &= \nabla_{\vec{\Phi}} \nabla_{\vec{\Phi}}^{T} V^{(0)} 
  = \nabla_{\vec{\Phi}} (4 \vec{\Phi} \circ \mat{\Lambda} \vec{\Phi}^{\circ 2})^{T}
  \\
  &= 4 \diag (\mat{\Lambda} \vec{\Phi}^{\circ 2} ) + 8 \mat{\Lambda} \circ (\vec{\Phi} \vec{\Phi}^{T}),
\end{split}
  \label{eq:m:sq:tree:biq}
\end{equation}
where $\diag(\vec{v})$ designates a diagonal matrix with entries given by the vector $\vec{v}$. For the first term, we have used $\nabla_{\vec{\Phi}} \vec{\Phi}^{T} = \mat{I}$.

\section{Effective potential}
\label{sec:V:eff}

Considering one-loop quantum corrections, the effective potential can be written as
\begin{equation}
  V = V^{(0)} + V^{(1)},
\label{eq:V:eff}
\end{equation}
where $V^{(0)}$ is the tree-level potential and $V^{(1)}$ is the radiative correction.

Using dimensional regularisation in the $\overline{\text{MS}}$ scheme, the one-loop contribution is given by
\begin{equation}
    V^{(1)} = \frac{1}{64 \pi^{2}} \Str \mat{m}^{4} 
    \left(\ln \frac{\mat{m}^{2}}{\mu^{2}} - \mat{C} \right),
\label{eq:V:(1)}
\end{equation}
where $\mu$ is the renormalisation scale. The supertrace ranges over tree-level field-dependent scalar, vector and fermion mass matrices $\mat{m}^{2}_{S,V,F}$. The matrix $\mat{C}$, peculiar to the $\overline{\text{MS}}$ scheme, is a block-diagonal constant matrix with diagonal elements equal to $c_{S,F} = \frac{3}{2}$ for scalars and fermions, and $c_{V} = \frac{5}{6}$ for gauge bosons.

The one-loop correction~\eqref{eq:V:(1)} can also be rewritten in the form \cite{Chataignier:2018aud}
\begin{equation}
    V^{(1)} = \mathbb{A} + \mathbb{B} \ln \frac{\varphi^{2}}{\mu^{2}},
\label{eq:V:(1):A:B}
\end{equation}
where
\begin{align}
  \mathbb{A} &= \frac{1}{64 \pi^{2}} \Str \mat{m}^{4} \left(\ln \frac{\mat{m}^{2}}{\varphi^{2}} - \mat{C} \right) = \frac{1}{64 \pi^{2}} \left\{ 
  \tr \left[ \mat{m}_{S}^{4} \left( \ln \frac{\mat{m}_{S}^{2}}{\varphi^{2}} - \frac{3}{2} \right) \right] 
 \right.
 \notag
 \\
 & \left.
  + 3 \tr \left[ \mat{m}_{V}^{4} \left( \ln \frac{\mat{m}_{V}^{2}}{\varphi^{2}} - \frac{5}{6} \right) \right]
  - 4 \kappa \tr \left[ \mat{m}_{F}^{4} \left( \ln \frac{\mat{m}_{F}^{2}}{\varphi^{2}} - \frac{3}{2} \right) \right]
  \right\},
\label{eq:A}
\\
 \mathbb{B} &= \frac{1}{64 \pi^{2}} \Str \mat{m}^{4} 
 = \frac{1}{64 \pi^{2}} \left( \tr \mat{m}_{S}^{4} + 3 \tr \mat{m}_{V}^{4} - 4 \kappa \tr \mat{m}_{F}^{4} \right),
\label{eq:B}
\end{align}
$\kappa = 1$ $(1/2)$ for Dirac (Weyl) fermions and $\varphi$ is the radial coordinate in the field space defined in eq.~\eqref{eq:Phi:norm}. Although in eqs.~\eqref{eq:V:(1):A:B}, \eqref{eq:A} and~\eqref{eq:B} one could use another pivot scale instead of $\varphi^{2}$, the radial coordinate is the most natural choice \cite{Chataignier:2018aud}. By construction, the scalar and vector boson terms in eq.~\eqref{eq:B} are always positive, while the fermion term is negative.

Notice that when computed at the minimum of the potential, the tree-level mass of the scale-invariance Goldstone is negative and results in a spurious imaginary part of the logarithm. The exact solution to this problem requires a shift of the Goldstone mass $m^{2}_{G} \to m^{2}_{G} + \Delta$ \cite{Elias-Miro:2014pca,Martin:2014bca} (see also e.g. \cite{Braathen:2016cqe}), but the numerical impact of the spurious part --- in particular at the one-loop level --- is negligible. In practice, it is then sufficient to consider only the real part of the logarithm, which corresponds to taking absolute values of the eigenvalues of the scalar mass matrix. Regardless, we will also show how to implement the exact solution in our formalism.

\subsection{Renormalisation group}
\label{subsec:rg}

The renormalisation scale $\mu$ is arbitrary and the effective potential must not depend on it. Therefore, changes in $\mu$ are to be counterbalanced by changes in the masses, couplings and fields, as formalised in the Callan-Symanzik equation
\begin{equation}
  \left( \mu \frac{\partial}{\partial \mu} + \sum_{i} \beta_{i} \frac{\partial}{\partial g_{i}} - \vec{\Phi}^{T} \mat{\gamma} \nabla_{\vec{\Phi}} \right) V = 0,
\label{eq:C-Z}
\end{equation}
where $g_{i}$ include all mass parameters and coupling constants and $\mat{\gamma}$ is the matrix of anomalous dimensions of the scalar fields. The functions $\beta_{i} = \mu \, d g_{i}/d \mu$ describe the running of couplings and masses with renormalisation scale.

In the case of a biquadratic tree-level potential, the $\beta$-function of the coupling matrix $\mat{\Lambda}$ is given by a matrix $\mat{\beta}$. The Callan-Symanzik equation \eqref{eq:C-Z} implies that $\mathbb{B}$, $\mat{\beta}$ and $\mat{\gamma}$ are related to each other via
\begin{equation}
  2 \mathbb{B} = (\vec{\Phi}^{\circ 2})^{T} \! \mat{\beta} \vec{\Phi}^{\circ 2} 
  - \vec{\Phi}^{T} \mat{\gamma} \nabla_{\vec{\Phi}} V^{(0)},
\label{eq:2:B:beta}
\end{equation}
where $\nabla_{\vec{\Phi}} V^{(0)} =  4 \vec{\Phi} \circ \mat{\Lambda} \vec{\Phi}^{\circ 2}$. At one-loop level, $\mat{\gamma} = \mat{0}$ in the absence of gauge and Yukawa interactions.  As shown in appendix~\ref{sec:beta:matrix},  the matrix $\beta$-function is given by
\begin{equation}
\begin{split}
  16 \pi^{2} \mat{\beta} &= 32 \mat{\Lambda}^{\circ 2} + 8 \mat{\Lambda}^{2}
  + 16 \mat{\Lambda} \Diag (\mat{\Lambda}) + 16 \Diag (\mat{\Lambda}) \mat{\Lambda} 
  \\
  & + \text{gauge and Yukawa contributions},
\end{split}
\label{eq:beta}
\end{equation} 
where $\Diag(\mat{\Lambda}) \equiv \mat{\Lambda} \circ \mat{I}$ is the diagonal matrix with the same diagonal as $\mat{\Lambda}$.
Conditions for Yukawa interactions to preserve the symmetries of biquadratic potentials are discussed in appendix~\ref{sec:fermion:mass:biquadratic} and a brief summary of gauge contributions is given in appendix~\ref{sec:gauge:biquadratic}. Our result \eqref{eq:beta} for biquadratic potentials agrees with general results in the literature \cite{Machacek:1983tz,Machacek:1983fi,Machacek:1984zw,Luo:2002ti}.

\subsection{Improved potential}
\label{subsec:impr:pot}

The validity of the perturbative expansion can be improved by taking a different renormalisation scale at each point of the field space. A common  choice to RG-improve the effective potential sets $\mu = \varphi$, or $t = \ln (\varphi/\varphi_{0})$, which cancels only the second term in the one-loop radiative correction~\eqref{eq:V:(1):A:B}, yielding%
\footnote{Notice that we could absorb the last term in~\eqref{eq:V:eff:mu:varphi} into the quartic coupling matrix by writing
$\mat{\Lambda}_{\text{eff}} \equiv \mat{\Lambda} + (\mathbb{A}/\varphi^{4}) \, \vec{e} \vec{e}^{T}$,
where we have used $(\vec{\Phi}^{\circ 2})^{T} \vec{e} \vec{e}^{T} \vec{\Phi}^{\circ 2} = \varphi^{4}$. That is, all quartic couplings are shifted by $\mathbb{A}/\varphi^{4}$. The last term can also be absorbed into the quartic coupling tensor in a general potential~\eqref{eq:pot:gen}, but contrary to the biquadratic case, no unique prescription can be given.}
\begin{equation}
    V = V^{(0)} + \mathbb{A}.
  \label{eq:V:eff:mu:varphi}
\end{equation}

Following the analysis in~\cite{Chataignier:2018aud}, we instead choose a field-dependent renormalisation scale $\mu_{*}$ such that the one-loop part $V^{(1)}$ of the effective potential~\eqref{eq:V:eff} vanishes (see appendix~\ref{sec:impro} for improvement procedures that use a different renormalisation scale). The effective potential then acquires the form of the tree-level potential, 
\begin{equation}
    V = V^{(0)}(t_*) = (\vec{\Phi}^{\circ 2})^T \! \mat{\Lambda}(t_*) \vec{\Phi}^{\circ 2},
  \label{eq:V:eff:t:tree}
\end{equation}
with running couplings in $\mat{\Lambda}(t_*)$ given by the solutions to the $\beta$-functions. The variable $t_*$ quantifies the displacement along the RG flow from the initial conditions set at a scale $\mu$ and is formally given by
\begin{equation}
  t_* = \frac{V^{(1)}}{2 \mathbb{B}}
  \label{t:tree:def}
\end{equation}
or, through eq.~\eqref{eq:V:(1):A:B}, by
\begin{equation}
  t_* = \frac{1}{2} \ln \frac{\varphi^{2}}{\mu^{2}} + \frac{1}{2} \frac{\mathbb{A}}{\mathbb{B}}.
\label{t:tree}
\end{equation}
An expression for $t_*$ can be found by solving $V^{(1)} = 0$ numerically at a given point of field space, yielding
\begin{equation}
  \mu_{*} = \mu e^{t_*}.
\end{equation}
 To a first approximation, $t_*$ can be written as  
\begin{equation}
  t_*^{(0)}  = \frac{1}{2} \ln \frac{\varphi^{2}}{\mu^{2}} + \frac{1}{2} \frac{\mathbb{A}_{0}}{\mathbb{B}_{0}},
\label{t:tree2}
\end{equation}
where the subscript in $\mathbb{A}_{0}$ and $\mathbb{B}_{0}$ indicates that the quantity is evaluated for couplings held at the fixed scale $\mu$, neglecting the running.
Because both $\mathbb{A}_{0}$ and $\mathbb{B}_{0}$ are homogeneous functions of order four, the first term  in \eqref{t:tree2}  depends only on the field vector length $\varphi$, while the second term is purely a function of the direction of the field vector. This distinction holds only at the level of $t_*^{(0)}$, since the second term in $t_*$ also depends on $\varphi$ through the running couplings. 

With the $t_*^{(0)}$ approximation, the method ceases to be valid on the $\mathbb{B}_{0}=0$ hypersurface which, being tangent to the RG flow, is a characteristic hypersurface of the RG equations ~\cite{Chataignier:2018aud}. Conversely, with $t_*$, the $\mathbb{B}=0$ hypersurface is not problematic because the running of couplings in $\mathbb{B}$ regularises the divergence that occurs in the approximate expression. In order to maintain analytic control, in the following we employ the approximation --- denoting henceforth $t_*^{(0)}$ with $t$ for simplicity ---  taking care of defining $t_*^{(0)}$ in a subspace where $\mathbb{B}_{0}\neq0$. In particular, after solving the RG equation for arbitrary initial conditions, we use the RG invariance of the effective potential to initialise the parameter in the vicinity of the radiative minimum, where $\mathbb{B}_{0} > 0$ holds as we show in section~\ref{subsec:mass:matrix}. Since $\mathbb{B}_{0}$ depends on the field values but does \emph{not} depend on the running couplings, this quantity necessarily remains positive everywhere but at the origin, the denominator in $t$ is finite and the RG flow well defined. 

Notice that $t_*^{(0)}$ coincides with $t_*$ at the scale $\mu$ where these parameters are defined. Consequently, with $t_*^{(0)}$ specified as above,   the approximation holds in the vicinity of the radiative minimum as the quantum corrections in $V^{(1)}$ necessarily remain small in the field space region of interest. Although this choice guarantees the precision required for the computation of the mass spectrum of the theory, we observe that the $t_*^{(0)}$ approximation progressively worsens when considering field values away from the radiative minimum.

The formalism of Ref.~\cite{Chataignier:2018aud} can also be applied beyond the one-loop order, though the analytical expression of $t_{*}$ becomes rather involved. At two-loop order~\cite{Martin:2001vx}, it is possible to obtain an approximate relation for $t_*$ by retaining the leading terms in the form $V^{(1)} + V^{(2)} \approx \mathbb{A} + \mathbb{B} \ln (\varphi^2/\mu^2) + \mathbb{C} [\ln (\varphi^2/\mu^2)]^{2}$.

\subsection{Anomalous dimensions}
\label{subsec:anom:dim}

If the matrix $\mat{\gamma}$ of anomalous dimensions does not vanish, the scalar fields are rescaled as $\vec{\Phi}(t) = e^{\mat{\Gamma}(t)} \vec{\Phi}(0)$, where 
\begin{equation}
  \mat{\Gamma}(t) = -\int_{0}^{t} \mat{\gamma}(s) ds,
\end{equation}
so $d\mat{\Gamma}(t)/dt = -\mat{\gamma}(t)$. In general, $\mat{\gamma}$ can be non-diagonal and may not commute with $e^{\mat{\Gamma}}$, but as a result of the $\mathbb{Z}_{2}^{n}$ symmetry of a biquadratic potential both $\mat{\gamma}$ and $e^{\mat{\Gamma}}$ are diagonal and commute.

Our effective potential then takes the form
\begin{equation}
  V = \left[(e^{\mat{\Gamma}} \vec{\Phi})^{\circ 2} \right]^{T} \! \mat{\Lambda} (e^{\mat{\Gamma}} \vec{\Phi})^{\circ 2}
  = (\vec{\Phi}^{\circ 2})^{T} e^{2 \mat{\Gamma}} \!  \mat{\Lambda}  e^{2 \mat{\Gamma}}\vec{\Phi}^{\circ 2},
  \label{eq:V:eff:t:tree:anom:dim}
\end{equation}
where we used the fact that $e^{\mat{\Gamma}}$ is diagonal to pull it through $\vec{\Phi}$.

If we ignore higher order terms involving two derivatives with respect to $t$, the formulae for the stationary point equation and the mass matrix, computed in the next section, will hold through the substitutions $\mat{\Lambda} \to e^{2 \mat{\Gamma}} \! \mat{\Lambda}  e^{2 \mat{\Gamma}}$ and $\mat{\beta} \to e^{2 \mat{\Gamma}} \! \mat{\beta}  e^{2 \mat{\Gamma}}$.

\section{Minima of the effective potential}
\label{sec:V:eff:min}

In this section, we first analyse the vacuum stability of the potential as it is a necessary condition to have a global minimum at finite field values. Then, we derive the stationary point equation, calculate the mass matrix and show how to solve the equation iteratively. Most of the discussion pertains to biquadratic potentials, but several results can be generalised to arbitrary cases by using tensor algebra.

\subsection{Vacuum stability}
\label{subsec:BfB}

A necessary condition for a finite global minimum to exist is that the scalar potential must be bounded from below, otherwise the theory is unstable and no state of lowest energy exists. We thus begin our investigation by requiring the stability of the scalar potential. 

Because the second term in eq.~\eqref{t:tree} depends only on the direction of the field vector,  $t$ is a monotonic function of scalar fields in the limit of large field values \cite{Chataignier:2018aud}:
\begin{equation}
  t_{\varphi \to \infty} = \frac{1}{2} \ln \frac{\varphi^{2}}{\mu^{2}}.
\end{equation}
The one-loop RG-improved effective potential is thus bounded from below if it obeys the tree-level stability conditions for values of the running quartic couplings given at large scales. 

A biquadratic tree-level potential~\eqref{eq:pot:biq} is not directly a function of the field $\vec{\Phi}$, but rather of its Hadamard square $\vec{\Phi}^{\circ 2}$, which is a vector of non-negative elements. Moreover, the potential is a homogeneous function of the fields. For these reasons, a necessary and sufficient condition for the biquadratic tree-level potential~\eqref{eq:pot:biq} to be bounded from below is that the matrix $\mat{\Lambda}$ be \emph{copositive} \cite{Kannike:2012pe}. A matrix $\mat{A}$ is copositive if it is non-negative on vectors with non-negative elements \cite{Motzkin:1952aa}, that is $\vec{x}^{T} \mat{A} \vec{x} \geq 0, \; \forall \vec{x} \geq 0$.

Copositive matrices include the usual positive-semidefinite matrices, thus a positive-semidefinite $\mat{\Lambda}$ is sufficient, but not necessary, for the stability of the potential. Moreover, although $\mat{\Lambda}$ could be positive-semidefinite at a certain scale, this property is generally not preserved by the running of couplings. Indeed, even if the only contributions to the running of $\mat{\Lambda}$ arise from scalar interactions, we can regroup the terms in the $\beta$-function~\eqref{eq:beta} as
\begin{equation}
  16 \pi^{2} \mat{\beta} = 32 [\mat{\Lambda}^{\circ 2} - \Diag(\mat{\Lambda}^{\circ 2})] + 8 [\mat{\Lambda} + 2 \Diag (\mat{\Lambda})]^{2},
\end{equation} 
which is a sum of a matrix with non-negative elements and a positive-semidefinite matrix. Therefore, the running $\mat{\Lambda}$ remains copositive but may not be positive-semidefinite if its non-diagonal elements are large compared to the diagonal entries. 

The copositivity of a matrix can be ascertained, for example, via the Cottle-Habetler-Lemke (CHL) theorem \cite{Cottle1970295}: Suppose that the order $n-1$ principal submatrices of a real symmetric matrix $\mat{A}$ of order $n$ are copositive. Then $\mat{A}$ is copositive if and only if 
\begin{equation}
  \det (\mat{A}) \geq 0 \quad \lor \quad \text{some element(s) of } \adj (\mat{A}) < 0.
\label{eq:CHL:cop}
\end{equation}
In other words, $\mat{A}$ is \textsc{not} copositive if and only if
\begin{equation}
  \det (\mat{A}) < 0 \quad \land \quad \text{all elements of } \adj (\mat{A}) \geq 0,
\label{eq:CHL:not:cop}
\end{equation}
that is, the inverse $\mat{A}^{-1}$ exists and its elements are all non-positive. The adjugate $ \adj(\mat{A})$ of a matrix $\mat{A}$ equals the transpose of the cofactor matrix of $\mat{A}$. It is implicitly defined through the relation $\mat{A} \adj(\mat{A}) = \det(\mat{A}) \, \mat{I}$.
Notice that $\det (\mat{\Lambda})$ can be negative even when the non-diagonal elements of $\mat{\Lambda}$ --- the portal couplings --- are large and positive, posing no danger for the potential to become unstable. In this case, some elements of the adjugate $\adj (\mat{\Lambda})$ are necessarily negative. 

It is now clear that the copositivity of the running coupling matrix $\mat{\Lambda}$ must hold in the limit of large field values for the effective potential to be bounded from below. On the contrary, we will show that a radiatively generated minimum requires, for its existence, a \emph{violation} of the copositivity conditions in a \emph{finite} range of field values. This criterion allows one to survey for the presence of an effective potential minimum solely on the basis of the running of quartic couplings. In particular, as we will see in section \ref{subsec:mass:matrix}, the potential $V < 0$ in a minimum, implying $\det(\mat{\Lambda}) < 0$. Therefore, by the theorem~\eqref{eq:CHL:not:cop}, all elements of $\adj (\mat{\Lambda})$ must be non-negative in order for the minimum to exist.

\subsection{Stationary point equation}
\label{subsec:min:eq}

We obtain the stationary point equation for the effective potential~\eqref{eq:V:eff:t:tree} by setting to zero its gradient with respect to the field vector $\vec{\Phi}$. Considering a biquadratic tree-level potential~\eqref{eq:pot:biq}, the equation for the stationary point of the effective potential is given by
\begin{equation}
\begin{split}
  \vec{0} = \nabla_{\vec{\Phi}} V&= 4 \vec{\Phi} \circ \mat{\Lambda} \vec{\Phi}^{\circ 2} 
  + (\vec{\Phi}^{\circ 2})^{T} \! \mat{\beta} \vec{\Phi}^{\circ 2} \, \nabla_{\vec{\Phi}} t 
  \\
  &= 4 \vec{\Phi} \circ \mat{\Lambda} \vec{\Phi}^{\circ 2} 
    + 2 \mathbb{B} \, \nabla_{\vec{\Phi}} t,
\end{split}
\label{eq:min:biq:t:tree}
\end{equation}
where we have used the definition of the matrix $\beta$-function $ \mat{\beta}=d \mat{\Lambda}/d t$, the chain rule and the relation \eqref{eq:2:B:beta}. As anticipated, we neglect anomalous dimensions which can easily be taken into account with the substitutions $\mat{\Lambda} \to e^{2 \mat{\Gamma}} \! \mat{\Lambda}  e^{2 \mat{\Gamma}}$ and $\mat{\beta} \to e^{2 \mat{\Gamma}} \! \mat{\beta}  e^{2 \mat{\Gamma}}$ in the results below. In the rest, $\vec{\Phi}$ will mostly denote the field vector in the minimum, i.e. the vector of VEVs. The VEV of the radial direction $\varphi=\sqrt{\vec{\Phi}^{T} \vec{\Phi}}$ is denoted by $v_{\varphi}$.

The field-space gradient of the $t$ variable is given by
\begin{equation}
  \nabla_{\vec{\Phi}} t = \frac{1}{\varphi^{2}} \vec{\Phi} + \frac{1}{2} \nabla_{\vec{\Phi}} \frac{\mathbb{A}_{0}}{\mathbb{B}_{0}},
   \label{eq:grad:t:tree}
\end{equation}
where we neglect terms proportional to $d\mathbb{A}/dt$ and $d\mathbb{B}/dt$  consistently with the adopted $t_*^{(0)}$ approximation. 

The two terms in the gradient~\eqref{eq:grad:t:tree} are orthogonal to each other. This is because $\mathbb{A}_{0}$ and $\mathbb{B}_{0}$ are both homogeneous functions of order four, so $\vec{\Phi}^{T} \nabla_{\vec{\Phi}} \mathbb{A}_{0} = 4 \mathbb{A}_{0}$ and $\vec{\Phi}^{T} \nabla_{\vec{\Phi}} \mathbb{B}_{0} = 4 \mathbb{B}_{0}$, due to which $\vec{\Phi}^{T} \nabla_{\vec{\Phi}} (\mathbb{A}_{0}/\mathbb{B}_{0}) = 0$ and, as a consequence, $\vec{\Phi}^{T} \nabla_{\vec{\Phi}} t  = 1$. Equivalently, we can consider that $\vec{\Phi}^{T} \nabla_{\vec{\Phi}} = \varphi \partial_{\varphi}$ \cite{AlexanderNunneley:2010nw}, which also leads to $\vec{\Phi}^{T} (\nabla_{\vec{\Phi}} \nabla_{\vec{\Phi}}^{T} t) \vec{\Phi} = -1$. 

Multiplying the stationary point equation~\eqref{eq:min:biq:t:tree} from the left by $\vec{\Phi}^{T}$ and using~\eqref{eq:2:B:beta}, we obtain the radial stationary point equation
\begin{equation}
\begin{split}
  0 = \vec{\Phi}^{T} \nabla_{\vec{\Phi}} V &= 4 (\vec{\Phi}^{\circ 2})^{T} \! \mat{\Lambda} \vec{\Phi}^{\circ 2} 
  + (\vec{\Phi}^{\circ 2})^{T} \! \mat{\beta} \vec{\Phi}^{\circ 2} \, \vec{\Phi}^{T}\nabla_{\vec{\Phi}} t 
  \\
  &= 4 V +2 \mathbb{B},
\end{split}
  \label{eq:min:biq:t:tree:radial}
\end{equation}
which, in the case of a single scalar field with self-coupling $\lambda$ and  $\beta$-function $\beta$, reduces to the well-known relation $4 \lambda + \beta = 0$ \cite{Yamagishi:1982hy,Einhorn:1982pp,Einhorn:2007rv}. Notice that the radial equation alone does not fully specify the minimum: it defines a $(n-1)$-dimensional hypersurface in the field space, but the minimum direction on this surface is determined only by eq.~\eqref{eq:min:biq:t:tree}. 

In the next subsection we show that $\mathbb{B} > 0$ in a minimum, hence for eq.~\eqref{eq:min:biq:t:tree:radial} we must have $V < 0$ at the same point in field space.\footnote{There is another solution $V = \mathbb{B} = 0$. Because in this case we must consider $d\mathbb{B}/dt$, it requires a two-loop level calculation. Notice that $V \approx 0$ can account for the observed tiny positive vacuum energy. We ignore this problem in the usual one-loop effective potential minimum with $V < 0$, which gives a too large cosmological constant of the wrong sign.}
Because the potential has to be bounded from below for large field values, it also follows that the potential vanishes at a radial distance beyond the minimum radius $v_{\varphi}$. Such a point, together with the origin of the field space, identifies the (tree-level) flat direction associated with the minimum, be it of a Coleman-Weinberg or Gildener-Weinberg type \cite{Chataignier:2018aud}.

The quantities $\mathbb{A}$ and $\mathbb{B}$ embodying the quantum corrections enjoy the same discrete symmetry as the biquadratic tree-level potential and, therefore, depend on scalar fields only via $\vec{\Phi}^{\circ 2}$. We then apply the chain rule,
\begin{equation}
  \nabla_{\vec{\Phi}} t = 2 \vec{\Phi} \circ \nabla_{\vec{\Phi}^{\circ 2}} t
\label{eq:chain}
\end{equation}
and write the stationary point equation~\eqref{eq:min:biq:t:tree} as 
\begin{equation}
\begin{split}
  0 &= 4 \vec{\Phi} \circ \mat{\Lambda} \vec{\Phi}^{\circ 2} + 4 \mathbb{B} \vec{\Phi} \circ \nabla_{\vec{\Phi}^{\circ 2}} t
  \\
  & = 4 \vec{\Phi} \circ \left( \mat{\Lambda} \vec{\Phi}^{\circ 2} + \mathbb{B}  \nabla_{\vec{\Phi}^{\circ 2}} t \right).
\end{split}
  \label{eq:min:biq:t:tree:via:radial}
\end{equation}
The first term in the resulting Hadamard product, $\vec{\Phi}$, gives the trivial vacuum solution: for a non-zero minimum, it can be used to force up to $n - 1$ VEVs to vanish.

\subsection{Non-trivial minima}
\label{subsec:min:ntm}

Let us assume, to begin with, that all of the VEVs are finite. Then in~\eqref{eq:min:biq:t:tree:via:radial} the factor in parentheses  must vanish. Using the radial equation~\eqref{eq:min:biq:t:tree:radial} to trade $\mathbb{B}$ for $V$, we have
\begin{equation}
  \mat{\Lambda} \vec{\Phi}^{\circ 2} = 2 V \nabla_{\vec{\Phi}^{\circ 2}} t.
  \label{eq:min:biq:t:tree:nonzero}
\end{equation}
Upon multiplying eq.~\eqref{eq:min:biq:t:tree:nonzero} from the left by $ \adj (\mat{\Lambda})$, we obtain
\begin{equation}
  \det (\mat{\Lambda}) \vec{\Phi}^{\circ 2} = 2 V \! \adj (\mat{\Lambda}) \nabla_{\vec{\Phi}^{\circ 2}} t,
  \label{eq:min:biq:t:tree:nonzero:adj}
\end{equation}
and observe that the potential value in the minimum is proportional to $\det (\mat{\Lambda})$. Using 
\begin{equation}
  \nabla_{\vec{\Phi}^{\circ 2}} t = \frac{1}{2} \left[ \frac{1}{v_{\varphi}^{2}} \vec{e} + \nabla_{\vec{\Phi}^{\circ 2}} \frac{\mathbb{A}_{0}}{\mathbb{B}_{0}} \right],
   \label{eq:grad:2:t:tree}
\end{equation}
we finally find an implicit solution for the minimum,
\begin{equation}
  \vec{\Phi}^{\circ 2} = \frac{V}{\det (\mat{\Lambda})} \left[ \frac{1}{v_{\varphi}^{2}} \adj (\mat{\Lambda}) \vec{e} + \adj (\mat{\Lambda}) \nabla_{\vec{\Phi}^{\circ 2}} \frac{\mathbb{A}_{0}}{\mathbb{B}_{0}} \right] \geq 0,
\label{eq:min:biq:t:tree:nonzero:final}
\end{equation}
where the inequality must be satisfied in order for the solution to be physical (with the equality as a limiting case). In section~\ref{subsec:sol:it} we solve eq.~\eqref{eq:min:biq:t:tree:nonzero:final} iteratively, starting from the lowest order solution presented in section~\ref{subsec:sol:0th:order}. 

The general expressions for $\nabla_{\vec{\Phi}^{\circ 2}} \mathbb{A}_{0}$ and $\nabla_{\vec{\Phi}^{\circ 2}} \mathbb{B}_{0}$ entering the second term of~\eqref{eq:min:biq:t:tree:nonzero:final} are 
\begin{align}
  \nabla_{\vec{\Phi}^{\circ 2}} \mathbb{B}_{0} &= \frac{1}{32 \pi^{2}} \Str \mat{m}^{2} \nabla_{\vec{\Phi}^{\circ 2}} \mat{m}^{2} = \mat{\beta} \vec{\Phi}^{\circ 2} 
  - 2 \mat{\gamma} \mat{\Lambda} \vec{\Phi}^{\circ 2} - 2  \mat{\Lambda} \mat{\gamma} \vec{\Phi}^{\circ 2},
  \\
  \nabla_{\vec{\Phi}^{\circ 2}} \mathbb{A}_{0} &= \frac{1}{64 \pi^{2}} \Str \left[ 2 \mat{m}^{2} \nabla_{\vec{\Phi}^{\circ 2}}  \mat{m}^{2} \, \left(\ln \frac{\mat{m}^{2}}{\varphi^{2}} - \mat{C} \right) + \mat{m}^{2} \nabla_{\vec{\Phi}^{\circ 2}}  \mat{m}^{2} - \mat{m}^{4} \frac{1}{\varphi^{2}} \vec{e}
   \right]
   \notag
   \\
 &= \frac{1}{64 \pi^{2}}  \Str \left( 2 \mat{m}^{2} \nabla_{\vec{\Phi}^{\circ 2}}  \mat{m}^{2} \,\ln \frac{\mat{m}^{2}}{\varphi^{2}}\right)
  + 
 \notag
  \frac{1}{2}  \nabla_{\vec{\Phi}^{\circ 2}} \mathbb{B}_{0}
  \\
   &- \sum_{i = S,F,V} c_{i} \nabla_{\vec{\Phi}^{\circ 2}} \mathbb{B}_{0,i}
  - 
 \mathbb{B}_{0} \frac{1}{\varphi^{2}} \vec{e},
 \label{eq:nabla:A:0}
\end{align}
where we used eqs.~\eqref{eq:B} and~\eqref{eq:2:B:beta}. The gradient of the mass matrix is more precisely written as $\nabla_{\vec{\Phi}^{\circ 2}} \otimes \mat{m}^{2}$, but we leave the tensor product implicit everywhere in order to avoid clutter. We denote with $\mathbb{B}_{S,F,V}$ the contribution of scalars, fermions and vectors to $\mathbb{B}$, respectively. The first term in second line of eq.~\eqref{eq:nabla:A:0} requires further action and, for purely scalar contributions, reduces to
\begin{equation}
\begin{split}
  \Str \left( 2 \mat{m}^{2}_{S} \nabla_{\vec{\Phi}^{\circ 2}}  \mat{m}^{2}_{S} \,\ln \frac{\mat{m}^{2}_{S}}{\varphi^{2}}\right) = 8 & \left[\mat{\Lambda} \left(\mat{m}^2_{S} \circ \ln \frac{\mat{m}^2_{S}}{\varphi^2}\right) \right] \vec{e} 
  \\
  + 16 & \left[\left(\mat{m}^2_{S} \ln \frac{\mat{m}^2_{S}}{\varphi^2} \right)\circ\mat{\Lambda} \circ (\vec{\Phi}^{\circ -1} \vec{e}^{T}) \right] \vec{\Phi}.
\end{split}
\label{eq:first:term:grad:A}
\end{equation}
While $\nabla_{\vec{\Phi}^{\circ 2}} \mathbb{B}_{0}$ can be written as the product of a matrix times the vector $\vec{\Phi}^{\circ 2}$, the gradient $\nabla_{\vec{\Phi}^{\circ 2}} \mathbb{A}_{0}$ cannot be written in this form. Therefore, it is impossible to solve the minimisation equation \eqref{eq:min:biq:t:tree:nonzero:final} analytically.

Notice that in the presence of a dominant mass scale $m^{2}$, eq.~\eqref{t:tree:def} gives 
\begin{equation}
  \nabla_{\vec{\Phi}^{\circ 2}} t \approx \frac{1}{2 m^{2}} \nabla_{\vec{\Phi}^{\circ 2}} m^{2},
\end{equation}
and also
\begin{equation}
  \nabla_{\vec{\Phi}^{\circ 2}} \nabla_{\vec{\Phi}^{\circ 2}}^{T} t 
  \approx \frac{1}{2 m^{2}} \left( - \frac{1}{m^{2}} \nabla_{\vec{\Phi}^{\circ 2}} m^{2} \nabla_{\vec{\Phi}^{\circ 2}}^{T} m^{2} + \nabla_{\vec{\Phi}^{\circ 2}} \nabla_{\vec{\Phi}^{\circ 2}}^{T} m^{2} \right).
\end{equation}

\subsection{Mass matrix and minimum conditions}
\label{subsec:mass:matrix}

A stationary point of the potential is a minimum if the eigenvalues of the associated mass matrix (the Hessian matrix) are positive. The full quantum-corrected mass matrix is given by
\begin{equation}
  \mat{M}^{2}_{S} = \nabla_{\vec{\Phi}} \nabla_{\vec{\Phi}}^{T} V
  = \mat{m}^{2}_{S} + 2 \nabla_{\vec{\Phi}} t \, \nabla_{\vec{\Phi}}^{T} \mathbb{B}
   + 2 \nabla_{\vec{\Phi}} \mathbb{B} \, \nabla_{\vec{\Phi}}^{T} t 
   + 2 \mathbb{B} \nabla_{\vec{\Phi}} \nabla_{\vec{\Phi}}^{T} t,
\label{eq:mass:matrix:full:biq}
\end{equation}
where the tree-level scalar mass matrix $\mat{m}^{2}_{S}$ is given in eq.~\eqref{eq:m:sq:tree:biq}, the gradient $\nabla_{\vec{\Phi}} t$ in eq.~\eqref{eq:grad:t:tree} and the gradient of $\mathbb{B}$ is 
\begin{equation}
  2 \nabla_{\vec{\Phi}} \mathbb{B} = 4 \vec{\Phi} \circ \mat{\beta} \vec{\Phi}^{\circ 2}
  - 8  \vec{\Phi} \circ \mat{\gamma} \mat{\Lambda} \vec{\Phi}^{\circ 2} - 8   \vec{\Phi} \circ  \mat{\Lambda} \mat{\gamma} \vec{\Phi}^{\circ 2},
  \label{eq:nabla:B:0}
\end{equation}
where we have neglected second-order terms which contain two derivatives of $t$ (such as $d \mat{\beta}/dt$). Finally, the double gradient of $t$ is given by
\begin{equation}
  \nabla_{\vec{\Phi}} \nabla_{\vec{\Phi}}^{T} t = \frac{1}{\varphi^{2}} \mat{I} -
  \frac{2}{\varphi^{4}} \vec{\Phi} \vec{\Phi}^{T} + \frac{1}{2} \nabla_{\vec{\Phi}} \nabla_{\vec{\Phi}}^{T} \frac{\mathbb{A}_{0}}{\mathbb{B}_{0}}.
  \label{eq:nabla:nabla:A:B}
\end{equation}
The explicit form of the last term of eq. \eqref{eq:nabla:nabla:A:B} is involved, but its most important properties can be readily seen. We know that $\vec{\Phi}^{T} (\nabla_{\vec{\Phi}} \nabla_{\vec{\Phi}}^{T} t) \vec{\Phi} = -1$ and it can be confirmed that indeed $\vec{\Phi}^{T} [\nabla_{\vec{\Phi}} \nabla_{\vec{\Phi}}^{T} (\mathbb{A}_{0}/\mathbb{B}_{0})] \vec{\Phi} = 0$. It can be shown, in addition, that the first two terms in eq.~\eqref{eq:nabla:nabla:A:B} are canceled by a part of the last term.

Sandwiching the mass matrix between the VEV vectors, we have
\begin{equation}
  \vec{\Phi}^{T} \mat{M}^{2}_{S} \vec{\Phi} = 8 \mathbb{B},
\label{eq:Phi:M:2:Phi:8:B}
\end{equation}
where we used $\vec{\Phi}^{T} \mat{m}^{2}_{S} \vec{\Phi} = 12 V =  -6 \mathbb{B}$ via the radial equation~\eqref{eq:min:biq:t:tree:radial}, $\vec{\Phi}^{T} \nabla_{\vec{\Phi}} \mathbb{B} = 4 \mathbb{B}$, and $\vec{\Phi}^{T} (\nabla_{\vec{\Phi}} \nabla_{\vec{\Phi}}^{T} t) \vec{\Phi} = -1$. 

Therefore $\mathbb{B} > 0$ is a necessary condition for the extremum to be a minimum (at least as long as couplings are perturbative and $d\mathbb{B}/dt$ is negligible). Consequently, the radial equation~\eqref{eq:min:biq:t:tree:via:radial} implies that the potential in the minimum respects $V < 0$. From $\vec{\Phi}^{T} \mat{m}^{2}_{S} \vec{\Phi} = 12 V$ we see that $V < 0$ is equivalent to having a negative eigenvalue in the tree-level mass matrix $\mat{m}^{2}_{S}$, and therefore $\mat{m}^{2}_{S} \prec \mat{0}$.

The potential can only be negative if the coupling matrix $\mat{\Lambda}$ is not copositive in the minimum. By the CHL theorem~\eqref{eq:CHL:not:cop}, we have that $\det (\mat{\Lambda}) < 0$ and all elements of $\adj (\mat{\Lambda})$ are non-negative. (All the submatrices of $\mat{\Lambda}$ are copositive, because we assumed that there are no minima in lower-dimensional subspaces of the field space.)
Thus, the running coupling matrix $\mat{\Lambda}$ has a negative eigenvalue.

Hence, in a minimum where the VEVs of all fields are non-zero, we have that
\begin{equation}
  \mathbb{B} > 0, \qquad V < 0, \qquad \det (\mat{\Lambda}) < 0 \land \adj (\mat{\Lambda}) \geq 0, \qquad \mat{m}^{2}_{S} \prec \mat{0}
\label{eq:min:conds}
\end{equation}
all hold and are equivalent.

Besides minima, quantum corrections may generate saddle points indicated by $V \geq 0$ (copositive $\mat{\Lambda}$) and $\mathbb{B} \leq 0$. If there is at least one finite minimum, then the origin is a saddle point with $V = \mathbb{B} = 0$. Therefore, if the potential has two local minima with a saddle point between them, then necessarily  $\mathbb{B} = 0$ at two points located between the saddle point and each minimum. A possible complication with the conditions will be discussed in the end of subsection~\ref{subsec:min:subspace}.

\subsection{Minimum solution to lowest order}
\label{subsec:sol:0th:order}

If we approximate $\nabla_{\vec{\Phi}^{\circ 2}} t$ by the first term in eq.~\eqref{eq:grad:2:t:tree}, then we have
\begin{equation}
  \mat{\Lambda} \vec{\Phi}^{\circ 2} = \frac{V}{v_{\varphi}^{2}} \, \vec{e}.
  \label{eq:min:biq:t:tree:almost:flat}
\end{equation}
The solution is given by
$\vec{\Phi}^{\circ 2} = v_{\varphi}^{2} \vec{n}^{\circ 2}$, where the unit vector in the direction of the minimum is
\begin{equation}
  \vec{n}^{\circ 2} = \frac{\adj (\mat{\Lambda}) \vec{e}}{\vec{e}^{T} \! \adj (\mat{\Lambda}) \vec{e}}.
\label{eq:min:biq:t:tree:almost:flat:sol}
\end{equation}
When $V = 0$, the vector $\vec{n}^{\circ 2}$ gives the tree-level flat direction. In the minimum, however, $V < 0$, so the lowest order solution $\vec{\Phi}^{\circ 2}$ is \emph{not} a null eigenvector of $\mat{\Lambda}$ evaluated at the minimum. However, for practical purposes, the difference between the minimum direction and the flat direction is negligible in most cases.\footnote{Because the matrix $\adj(\mat{\Lambda})$ has only positive elements, it is guaranteed to have a Perron-Frobenius eigenvector $\vec{v}$ with all elements positive, corresponding to the largest eigenvalue $r$ of the matrix. It is interesting that the vector \eqref{eq:min:biq:t:tree:almost:flat:sol}, normalised, is usually very close to the Perron-Frobenius eigenvector.  The eigenvalues of $\adj(\mat{\Lambda})$ are products of the eigenvalues of $\mat{\Lambda}$ with one eigenvalue absent in each product. At the flat direction, $\mat{\Lambda}$ has a zero eigenvalue, so $\adj(\mat{\Lambda})$ is a projection matrix with only one non-zero positive eigenvalue and the vector \eqref{eq:min:biq:t:tree:almost:flat:sol} \emph{is} the Perron-Frobenius eigenvector. In the minimum, the other eigenvalues of $\adj(\mat{\Lambda})$ are non-zero and negative, but small compared to the Perron-Frobenius eigenvalue, so $\adj (\mat{\Lambda}) \approx r \vec{v} \vec{v}^{T}$ and thus $\vec{n}^{\circ 2} \propto \vec{v}$.}

Substituting the solution~\eqref{eq:min:biq:t:tree:almost:flat:sol} into the radial minimisation equation~\eqref{eq:min:biq:t:tree:radial} yields
\begin{equation}
  0 = 4 \det (\mat{\Lambda}) \vec{e}^{T} \! \adj (\mat{\Lambda}) \vec{e}
  + \vec{e}^{T} \! \adj (\mat{\Lambda}) \mat{\beta} \adj (\mat{\Lambda}) \vec{e}
\end{equation}
or
\begin{equation}
  \det (\mat{\Lambda}) = -\frac{1}{4} \frac{\vec{e}^{T} \! \adj (\mat{\Lambda}) \mat{\beta} \adj (\mat{\Lambda}) \vec{e}}{\vec{e}^{T} \! \adj (\mat{\Lambda}) \vec{e}},
\label{eq:min:radial:zeroth}
\end{equation}
giving a condition on the running couplings that must hold in the minimum at the lowest order of approximation.

For the value of the effective potential at the minimum, we find  
\begin{equation}
  V= \frac{\det (\mat{\Lambda})}{\vec{e}^{T} \! \adj (\mat{\Lambda}) \vec{e}} v_{\varphi}^{4},
  \label{eq:V:lowest}
\end{equation}
whereas the quantum-corrected mass matrix, in this approximation, is 
\begin{equation}
  \mat{M}^{2}_{S} = \vec{\Phi} \vec{\Phi}^{T} \circ \left[8 \mat{\Lambda} 
  +  \frac{4}{v_{\varphi}^{2}} \Big( (\mat{\beta} \vec{\Phi}^{\circ 2}) \vec{e}^{T} +  \vec{e} (\mat{\beta} \vec{\Phi}^{\circ 2})^{T} - \frac{\mathbb{B}}{v_{\varphi}^{2}} \vec{e} \vec{e}^{T} \Big) \right].
\label{eq:M:zeroth}
\end{equation}
In fact, the last term in eq.~\eqref{eq:M:zeroth} can be dropped, because it will be canceled by a part of the $\nabla_{\vec{\Phi}} \nabla_{\vec{\Phi}}^{T} (\mathbb{A}_{0}/\mathbb{B}_{0})$ term (see discussion below eq.~\eqref{eq:nabla:nabla:A:B}).

We can now obtain the usual Gildener-Weinberg potential \cite{Gildener:1976ih} in the radial direction by making further approximations \cite{Sher:1988mj}. First, we insert the lowest order minimum solution~\eqref{eq:min:biq:t:tree:almost:flat:sol} into the potential~\eqref{eq:V:eff:mu:varphi}. Then, by approximating the running to the linear order in $t$, we have
\begin{equation}
\begin{split}
   V(\varphi) &= (\vec{n}^{\circ 2})^T \varphi^2 \mat{\Lambda}(\varphi) \, \varphi^2 \, \vec{n}^{\circ 2} + \mathbb{A}
   \\
   &= \varphi^4 (\vec{n}^{\circ 2})^T \left[\mat{\Lambda}(v_\varphi) 
   + \mat{\beta}(v_\varphi)  \ln \frac{\varphi}{v_\varphi} \right] \vec{n}^{\circ 2} 
   + \mathbb{A}(\vec{n}) \varphi^{4}
   \\
   &= \varphi^{4}  \left[ V^{(0)}(\vec{n}) + \mathbb{A}(\vec{n}) + \mathbb{B}(\vec{n}) \ln \frac{\varphi^2}{v_\varphi^2} \right]
   \\
   &= \mathbb{B}(\vec{n}) \varphi^{4} \left(\ln \frac{\varphi^2}{v_\varphi^2} - \frac{1}{2} \right),
\end{split}
\end{equation}
where we have used the radial equation~\eqref{eq:min:biq:t:tree:radial} in the last step.

\subsection{Iterative solution}
\label{subsec:sol:it}

In order to go beyond the Gildener-Weinberg approximation, we solve the minimum equation~\eqref{eq:min:biq:t:tree:nonzero:final} iteratively. After selecting an RG flow by giving initial conditions for the couplings at an arbitrary scale $\mu_{\rm in}$, we start the iteration procedure from eq.~\eqref{eq:min:biq:t:tree:almost:flat:sol},
\begin{equation}
  \vec{n}^{\circ 2}_{0} = \frac{\adj (\mat{\Lambda}_{0}) \vec{e}}{\vec{e}^{T} \! \adj (\mat{\Lambda}_{0}) \vec{e}},
\label{eq:min:biq:t:tree:0th}
\end{equation}
which gives the minimum direction at zeroth order. The zeroth order coupling matrix $\mat{\Lambda}_{0} = \mat{\Lambda} (t_{0})$ is obtained by inserting $\vec{n}^{\circ 2}_{0}$ in the radial equation~\eqref{eq:min:biq:t:tree:radial} --- resulting in eq.~\eqref{eq:min:radial:zeroth} --- and finding the scale $\mu_0=\mu_{\rm in}e^{t_{0}}$ at which the radial equation is satisfied. In order to improve the accuracy of the $t_*^{(0)}$ approximation, we now use the RGE solutions to reparametrise the running of couplings, taking as equivalent initial conditions their values at $\mu_0$. As a consequence, after the reparametrisation, eq.~\eqref{t:tree2} reads $t_0\to t(\varphi=v_{\varphi 0}, \mu=\mu_0)=0$ and we can solve it to obtain the norm $v_{\varphi 0}$ of the zeroth order solution $\vec{\Phi}^{\circ 2}_{0} = v_{\varphi 0}^{2} \vec{n}^{\circ 2}_{0}$.  

The first order solution is then obtained
by inserting $\vec{\Phi}^{\circ 2}_{0}$ into the RHS of eq.~\eqref{eq:min:biq:t:tree:nonzero:final}:
\begin{equation}
  \vec{\Phi}^{\circ 2}_{1} = \frac{V}{\det (\mat{\Lambda}_{1})} \left[ \frac{1}{v_{\varphi0}^{2}} \adj (\mat{\Lambda}_{1}) \vec{e} + \adj (\mat{\Lambda}_{1}) \nabla_{\vec{\Phi}^{\circ 2}} \frac{\mathbb{A}_{0}}{\mathbb{B}_{0}} \right]_{\vec{\Phi}^{\circ 2} = \vec{\Phi}^{\circ 2}_{0}},
\label{eq:sol:1st:order}
\end{equation}
where again $\mat{\Lambda}_{1} = \mat{\Lambda} (t_{1})$  satisfies the radial equation~\eqref{eq:min:biq:t:tree:radial} at the scale $\mu_1= \mu_0 e^{t_{1}}$. Proceeding as before, we obtain the first iteration $v_{\varphi 1}$ for the norm after reparametrising the RGE solutions so that $t_1 \to t (\varphi = v_{\varphi 1}, \mu=\mu_1)=0$. The procedure can be repeated until sufficient precision is attained.

Notice that, at first order, the second term in $\vec{\Phi}^{\circ 2}_{1}$ is orthogonal to $\vec{e}$:
\begin{equation}
\left. \vec{e}^{T} \adj (\mat{\Lambda}_{1}) \nabla_{\vec{\Phi}^{\circ 2}} \frac{\mathbb{A}_{0}}{\mathbb{B}_{0}} \right|_{\vec{\Phi}^{\circ 2} = \vec{\Phi}^{\circ 2}_{0}} = 0.
\end{equation}

The method can also implement the solution to the spurious negative mass of the Goldstone boson proposed in \cite{Elias-Miro:2014pca,Martin:2014bca}, by shifting --- at each iteration step --- the field-dependent scalar mass matrix $\mat{m}^{2}_{S}$ by $\Delta \, \vec{\Phi}_{G} \vec{\Phi}_{G}^{T}$ at that step, where $\Delta$ is the correction to the Goldstone mass and $\vec{\Phi}_{G}$ is the Goldstone direction at that order.

A criterion for the convergence of the iteration is that the spectral radius of the Jacobian --- the largest of absolute values of its eigenvalues --- of the RHS of eq.~\eqref{eq:min:biq:t:tree:nonzero:final} at the fixed point be less than unity. To assess convergence it is necessary to evaluate the Jacobian at least to the first order: the potentially large term $\nabla_{\vec{\Phi}^{\circ 2}} (\mathbb{A}_{0}/\mathbb{B}_{0})$, which could --- in principle --- spoil the convergence, does not enter at zeroth order. The Jacobian is calculated in detail in the appendix \ref{sec:convergence}. In the example analysed below, we find that the iteration converges to a fixed point for a large range of couplings.

\subsection{Inverse problem}
\label{subsec:inv:problem}

In the previous sections, after specifying the quartic, gauge and Yukawa couplings at a given scale, we have solved the minimisation equations and found the minimum of the potential together with the mass matrix. We consider now the opposite problem: given the full mass matrix $\mat{M}^{2}$ and the desired minimum $\vec{\Phi}^{\circ 2}$, is it possible to find a corresponding quartic coupling matrix $\mat{\Lambda}$ and the matrix $\beta$-function at that point?

In the mass matrix~\eqref{eq:mass:matrix:full:biq}, $\mat{\Lambda}$ appears explicitly in the tree-level part~\eqref{eq:m:sq:tree:biq}:
\begin{equation}
  \mat{m}^{2}_{S} = \diag ( 4 \mat{\Lambda} \vec{\Phi}^{\circ 2} ) + 8 \mat{\Lambda} \circ (\vec{\Phi} \vec{\Phi}^{T}).
\end{equation} In the first term, we can use the minimum condition~\eqref{eq:min:biq:t:tree} to substitute $4 \mat{\Lambda} \vec{\Phi}^{\circ 2} = -2 \mathbb{B} \vec{\Phi}^{\circ -1} \circ \nabla_{\vec{\Phi}} t$. Therefore,
\begin{equation}
\begin{split}
  \mat{\Lambda} &= \frac{1}{8} (\vec{\Phi} \vec{\Phi}^{T})^{\circ -1} \circ \left[ \mat{M}^{2}_{S} 
  + 2 \mathbb{B} \diag (\vec{\Phi}^{\circ -1} \circ \nabla_{\vec{\Phi}} t)
  \right. 
  \\
  &
  \left.
  - 2 \nabla_{\vec{\Phi}} t \, \nabla_{\vec{\Phi}}^{T} \mathbb{B} 
  - 2 \nabla_{\vec{\Phi}} \mathbb{B} \, \nabla_{\vec{\Phi}}^{T} t 
  - 2 \mathbb{B} \nabla_{\vec{\Phi}} \nabla_{\vec{\Phi}}^{T} t \right].
\end{split}
\label{eq:inv:Lambda}
\end{equation}
We can now approximate $t$ to the lowest order, $ t \approx (1/2) \ln (\varphi^{2}/\mu^{2})$, so that $\nabla_{\vec{\Phi}} t$ and $\nabla_{\vec{\Phi}} \nabla_{\vec{\Phi}}^{T} t$ depend on $\vec{\Phi}$ only. At this stage, with $8 \mathbb{B} = \vec{\Phi}^{T} \mat{M}_{S}^{2} \vec{\Phi}$, the only unknown is the gradient $2 \nabla_{\vec{\Phi}} \mathbb{B} = 4 \vec{\Phi} \circ \mat{\beta} \vec{\Phi}^{\circ 2}$. Although in principle the $\beta$-functions are all given in terms of $\mat{\Lambda}$, gauge and Yukawa couplings, still the dependence is non-linear.

We solve the issue by considering the lowest order solution~\eqref{eq:M:zeroth} for $\mat{M}^{2}_{S}$, where we neglect the dependence on the unknown $\nabla_{\vec{\Phi}} \mathbb{B}$ and drop the last term, thereby obtaining that
\begin{equation}
  \mat{\Lambda} \approx \frac{1}{8} \left(\vec{\Phi} \vec{\Phi}^{T} \right)^{\circ -1} \circ \mat{M}^{2}_{S}.
  \label{eq:inv:Lambda:0}
\end{equation}
The above formula can be used as a starting point for a ``reverse'' iteration that leads to the desired $\mat{\Lambda}$. At higher orders, we can use the expression \eqref{t:tree2} for $t$ and also obtain the minimum scale $\mu$ from $t(\varphi = v_{\varphi} = \sqrt{\vec{\Phi}^{T}\vec{\Phi}}, \mu) = 0$. 

\subsection{Minimum in a field subspace}
\label{subsec:min:subspace}

If up to $n - 1$ components of the field vector $\vec{\Phi}$ vanish in the minimum,  we can bring  the coupling matrix, the $\beta$-function matrix $\mat{\beta}$ and the field vector and $\nabla_{\vec{\Phi}} t$ into a block form upon a suitable permutation of the fields:
\begin{equation}
  \mat{\Lambda} = 
  \begin{pmatrix}
    \mat{\Lambda}_{11} & \mat{\Lambda}_{12}
    \\
    \mat{\Lambda}_{12}^{T} &\mat{\Lambda}_{22}
  \end{pmatrix},
  \quad
  \mat{\beta} = 
  \begin{pmatrix}
    \mat{\beta}_{11} & \mat{\beta}_{12}
    \\
    \mat{\beta}_{12}^{T} &\mat{\beta}_{22}
  \end{pmatrix},
  \quad
   \vec{\Phi}^{\circ 2} = 
  \begin{pmatrix}
    \vec{\Phi}^{\circ 2}_{1} \\ \vec{0}
  \end{pmatrix},
  \quad
  \nabla_{\vec{\Phi}} t = 
  \begin{pmatrix}
    \nabla_{\vec{\Phi}_{1}} t \\ \vec{0}
  \end{pmatrix}.
\label{eq:block}
\end{equation}
Consistency with the null VEVs forces the elements of $\nabla_{\vec{\Phi}} t$ corresponding to zero elements of $\vec{\Phi}$ to vanish too, otherwise these VEVs could be pushed to non-zero values. That the elements do vanish follows from the fact that the effective potential enjoys the same symmetries as the tree-level potential.

Therefore, we must solve the stationary point equations~\eqref{eq:min:biq:t:tree:nonzero} and~\eqref{eq:min:biq:t:tree:radial} restricted to the $\vec{\Phi}_{1}$ subspace:
\begin{align}
  \mat{\Lambda} \vec{\Phi}_{1}^{\circ 2} &= 2 V \nabla_{\vec{\Phi}_{1}^{\circ 2}} t,
  \label{eq:min:biq:t:tree:nonzero:subspace}
  \\
  0 &= 4 V + 2 \mathbb{B} =  4 (\vec{\Phi}_{1}^{\circ 2})^{T} \! \mat{\Lambda}_{11} \vec{\Phi}_{1}^{\circ 2} + (\vec{\Phi}_{1}^{\circ 2})^{T} \! \mat{\beta}_{11} \vec{\Phi}_{1}^{\circ 2},
  \label{eq:min:biq:t:tree:radial:subspace}
\end{align}
where the running of $\mat{\Lambda}_{11}$ still depends directly on $\mat{\Lambda}_{12}$ and, indirectly, on $\mat{\Lambda}_{22}$. 

The full mass matrix also admits a block form
\begin{equation}
   \mat{M}^{2}_{S} = 
  \begin{pmatrix}
    \mat{M}^{2}_{11} & \mat{M}^{2}_{12}
    \\
    (\mat{M}^{2}_{12})^{T} &\mat{M}^{2}_{22}
  \end{pmatrix}
  \quad
\end{equation}
with
\begin{align}
  \mat{M}^{2}_{11} &= 4 \diag(\mat{\Lambda}_{11} \vec{\Phi}_{1}^{\circ 2}) + 8 \mat{\Lambda}_{11}
  \circ (\vec{\Phi}_{1} \vec{\Phi}_{1}^{T}) 
  + 2 \nabla_{\vec{\Phi}_{1}} t \, \nabla_{\vec{\Phi}_{1}}^{T} \mathbb{B}
  + 2 \nabla_{\vec{\Phi}_{1}} \mathbb{B} \, \nabla_{\vec{\Phi}_{1}}^{T} t
  \notag
  \\
  &+ 2 \mathbb{B} \nabla_{\vec{\Phi}_{1}} \nabla_{\vec{\Phi}_{1}}^{T} t,
  \\
  \mat{M}^{2}_{22} &= 4 \diag(\mat{\Lambda}_{12}^{T} \vec{\Phi}_{1}^{\circ 2}) + 2 \mathbb{B} \nabla_{\vec{\Phi}_{2}} \nabla_{\vec{\Phi}_{2}}^{T} t,
  \\
    \mat{M}^{2}_{12} &= \mat{0},
\end{align}
where $\vec{\Phi}_{2}$ is the subspace of fields with null VEVs.
In fact, only the fields in the $\vec{\Phi}_{1}$ subspace, whose VEVs break the corresponding $\mathbb{Z}_{2}$ symmetries, are allowed to mix with each other and $\mat{M}^{2}_{22}$ then necessarily remains a diagonal matrix.

A positive-definite $\mat{M}^{2}_{22}$ constrains the off-diagonal $\mat{\Lambda}_{12}$, which also enters eqs.~\eqref{eq:min:biq:t:tree:nonzero:subspace} and~\eqref{eq:min:biq:t:tree:radial:subspace} via $\mat{\beta}_{11}$. The only direct constraint on $\mat{\Lambda}_{22}$ is that it be copositive.

As discussed in detail in section~\ref{sec:multi:min}, the effective potential can have more than one minimum. If two minima are located in the same field subspace (e.g. on the same axis), or in two orthogonal subspaces (e.g. two different axes), the corresponding existence conditions are independent of each other. It is also possible that two stationary points fulfil minimum conditions in part (e.g. have $V < 0$), while only one of them can be an actual minimum. Such a possibility arises when a stationary point lies in a subspace on the border of the subspace of the other point, for instance with one stationary point on the $\phi_{1}$-axis and another on the $\phi_{1} \phi_{2}$-plane. In such cases, it is possible to assess which stationary point is an actual minimum by checking if the associated mass matrix is positive-definite.

\subsection{Generalisation to generic potentials}
\label{subsec:general:pot}

In general, a scalar potential can have the form \eqref{eq:pot:gen} and, unlike a biquadratic potential \eqref{eq:pot:biq}, contain terms such as $\phi_{1}^{3} \phi_{2}$. While the minimum solution for the biquadratic case has a specific form, other results carry over with minimal changes. In tensor notation, we can write the RG-improved potential as
\begin{equation}
  V = \ten{\Lambda}(t) \vec{\Phi}^{4},
\label{eq:ten:pot}
\end{equation}
where $\ten{\Lambda}$ is the tensor of quartic couplings. A polynomial $f(\vec{x})$ of order  $m$ and its coefficient tensor $\ten{A}$ are related as $\ten{A} \vec{x}^{m} = m f(\vec{x})$ and $\ten{A} \vec{x}^{m-1} = \nabla_{\vec{x}} f(\vec{x})$ with
\begin{equation}
  \sum_{i_{1}, i_{2}, \ldots, i_{m}} A_{i_{1}i_{2}\ldots i_{m}} x_{i_{1}} x_{i_{2}} \cdots x_{i_{m}} = \vec{x}^{T} \! \ten{A} \vec{x}^{m-1}.
\end{equation}

The minimisation equations are given by
\begin{equation}
  \vec{0} = \nabla_{\vec{\Phi}} V = 4 \ten{\Lambda} \vec{\Phi}^{3} + \ten{\beta} \vec{\Phi}^{4} \, \nabla_{\vec{\Phi}} t,
  \label{eq:minim:tensor}
\end{equation}
where $\ten{\beta}$ is the $\beta$-function tensor. The radial equation is
\begin{equation}
  0 = 4 \ten{\Lambda} \vec{\Phi}^{4} + \ten{\beta} \vec{\Phi}^{4},
  \label{eq:min:V:beta:tensor}
\end{equation}
because $\vec{\Phi}^{T} \nabla_{\vec{\Phi}} (\mathbb{A}_{0}/\mathbb{B}_{0}) = 0$ due to the homogeneity of $\mathbb{A}_{0}$ and $\mathbb{B}_{0}$.

The full mass matrix is given by eq.~\eqref{eq:mass:matrix:full:biq} with
the scalar field-dependent mass matrix being
\begin{equation}
  \mat{m}^{2}_{S} = 12 \ten{\Lambda} \vec{\Phi}^{2},
\end{equation}
and the quantities
\begin{align}
  2 \mathbb{B} &= \ten{\beta} \vec{\Phi}^{4} - 4 \vec{\Phi}^{T} \mat{\gamma} \ten{\Lambda} \vec{\Phi}^{3},
  \\
  2 \nabla_{\vec{\Phi}} \mathbb{B} &= 4 \ten{\beta} \vec{\Phi}^{3} - 4 \mat{\gamma} \ten{\Lambda} \vec{\Phi}^{3} - 12 \vec{\Phi}^{T} \mat{\gamma} \ten{\Lambda} \vec{\Phi}^{2}.
\end{align}

Consequently, the relation $\vec{\Phi}^{T} \mat{M}^{2}_{S} \vec{\Phi} = 8 \mathbb{B}$ still holds and $\mathbb{B} > 0$ is a necessary condition for the minimum to exist. Therefore $V < 0$ in the minimum,  the $\ten{\Lambda}$ tensor must not be positive-definite and one of the \emph{tensor eigenvalues} $\lambda$ of $\ten{\Lambda}$ must be negative.

The \emph{E-eigenvalue equations} for the $\ten{\Lambda}$ tensor \cite{Qi20051302,2006math......7648L} are given by
\begin{align}
  \ten{\Lambda} \vec{\Phi}^{3} &= \lambda \vec{\Phi},
  \label{eq:Lambda:E:evals}
  \\
  \vec{\Phi}^{T} \vec{\Phi} &= 1,
  \label{eq:Lambda:E:sph}
\end{align}
but tensors also admit \emph{N-eigenvectors} and \emph{N-eigenvalues}, given by the solutions of $\ten{A} \vec{x}^{m-1} = \lambda \vec{x}^{\circ (m-1)}$. We refer the reader to ref. \cite{doi:10.1137/1.9781611974751} for a thorough discussion.

The number of E-eigenvectors of a symmetric tensor of order $m$ in $\mathbb{R}^n$ is
\begin{equation}
  d = \frac{(m - 1)^{n} - 1}{m - 2}.
\label{eq:number:eigenvectors}
\end{equation}
They can be complex, but only real tensor eigenvectors and eigenvalues are physical solutions of eqs.~\eqref{eq:Lambda:E:evals} and \eqref{eq:Lambda:E:sph}.

Eliminating $\phi_{i}$ from eqs.~\eqref{eq:Lambda:E:evals} and \eqref{eq:Lambda:E:sph} yields the characteristic polynomial $\phi_{\ten{\Lambda}}(\lambda)$ of the tensor, its degree is given by eq.~\eqref{eq:number:eigenvectors}. The multivariate resultant of a system of polynomial equations is a polynomial in their coefficients, which vanishes if and only if the equations have a common root. The free term of $\phi_{\ten{\Lambda}}(\lambda)$ --- the product of all E-eigenvalues --- is given by the resultant $\res_{\vec{\Phi}} (\ten{\Lambda} \vec{\Phi}^{3})$. For that reason, the resultant is also called the hyperdeterminant. Having a negative tensor eigenvalue then requires $\res_{\vec{\Phi}} (\ten{\Lambda} \vec{\Phi}^{3}) < 0$. 

In the limit of large field values, the tensor $\ten{\Lambda}$ must be positive-definite in order for the potential to be bounded from below: all of its eigenvalues and those of its principal subtensors --- obtained by setting one or more fields to zero --- must be positive \cite{Kannike:2016fmd,Ivanov:2018jmz}.

Note that for a quartic potential of two fields, the resultant is proportional to its discriminant computed with one field set to unity. Unfortunately, for a larger number of variables, the calculation of the resultant is computationally expensive, but the tensor eigenvalues can still be readily calculated numerically. Unlike for biquadratic potentials, in general all the potential coefficients cannot be determined by specifying the desired mass matrix.

\section{Examples}
\label{sec:examples}

In general, it is not possible to find analytically the direction in field space along which the minimum of the effective potential lies. Two notable exceptions are given by the case where a minimum is on an axis of field space and by the ``democratic'' setup, where the minimum lies equally distant from all axes. These configurations are, in a sense, complementary: the minimum on an axis is the extreme case of the minimum in a subspace (section~\ref{subsec:min:subspace}), where all but one fields are set to zero. In the democratic case, instead, all fields are equally non-zero. Despite this evident difference, the scenarios are united by the fact that the second term in $\nabla_{\vec{\Phi}} t$ identically vanishes and, therefore, the lowest order minimum solution~\eqref{eq:min:biq:t:tree:almost:flat:sol} is exact. To go beyond these simple examples, we will resort to the iterative solution of section \ref{subsec:sol:it} for more realistic cases.

In all the examples of this section, we use a simple two-field model with  scalar potential
\begin{equation}
  V = \lambda_{1} \phi_{1}^{4} + \lambda_{12} \phi_{1}^{2} \phi_{2}^{2} + \lambda_{2} \phi_{2}^{4}.
 \label{eq:2:field:V}
\end{equation}
Then, the scalar field vector is $\vec{\Phi} = ( \phi_{1} , \phi_{2} )^{T}$
and the quartic coupling matrix and its adjugate are, respectively, given by
\begin{equation}
  \mat{\Lambda} = 
  \begin{pmatrix} 
    \lambda_{1} & \frac{1}{2} \lambda_{12} \\
    \frac{1}{2} \lambda_{12} & \lambda_{2}
  \end{pmatrix},
  \qquad
    \adj(\mat{\Lambda})=
	\begin{pmatrix}
	\lambda_2 & -\frac{1}{2}\lambda_{12} \\
	-\frac{1}{2}\lambda_{12} & \lambda_1
  \end{pmatrix}.
\end{equation}
The tree-level field dependent masses are 
\begin{equation}
\begin{split}
  m^{2}_{\pm} &= (6 \lambda_{1} + \lambda_{12}) \phi_{1}^{2} + (6 \lambda_{2} + \lambda_{12}) \phi_{2}^{2} 
  \\
  &\pm \sqrt{(\lambda_{12} - 6 \lambda_{1})^{2} \phi_{1}^{4} +2 [7 \lambda_{12}^{2} + 6 \lambda_{12} (\lambda_{1} + \lambda_{2}) - 36 \lambda_{1} \lambda_{2}] \phi_{1}^{2} \phi_{2}^{2} + (\lambda_{12} - 6 \lambda_{2})^{2} \phi_{2}^{4}}.
\end{split}
\end{equation}
The quantities that enter the $t$ parameter are
\begin{equation}
  64 \pi^{2} \mathbb{B} = m_{-}^{4} + m_{+}^{4}
\end{equation}
and
\begin{equation}
  64 \pi^{2} V^{(1)} =  m^{4}_{-} \left(\ln \frac{\abs{m^{2}_{-}}}{\mu^{2}} - \frac{3}{2} \right) + m^{4}_{+} \left(\ln \frac{m^{2}_{+}}{\mu^{2}} - \frac{3}{2} \right),
\end{equation}
where we take the absolute value of the $m^{2}_{-}$ to deal with the spurious negative Goldstone mass.
The one-loop $\beta$-functions are given by
\begin{align}
  16 \pi^{2} \beta_{\lambda_{1}} &= 72 \lambda_{1}^{2} + 2 \lambda_{12}^{2},
  \\
  16 \pi^{2} \beta_{\lambda_{2}} &= 72 \lambda_{2}^{2} + 2 \lambda_{12}^{2},
  \\
  16 \pi^{2} \beta_{\lambda_{12}} &= 8 \lambda_{12} (3 \lambda_{1} + 3 \lambda_{2} + 2 \lambda_{12}).
\end{align}

\subsection{Minimum on an axis}
\label{subsec:min:axis}

We start with the simplest case, corresponding to a minimum of the effective potential that lies along a field space coordinate axis. This is also the simplest example of the possibilities analysed in section~\ref{subsec:min:subspace}, where not all the fields develop a finite VEV. 

Let the solution with a non-zero $v_{\varphi}$ be in the direction $\vec{\Phi} = v_{\varphi} \vec{e}_{i}$, where $\vec{e}_{i}$ is the unit vector along the $i$-th axis. Note that $\vec{e}_{i}^{\circ 2} = \vec{e}_{i}$. From the radial minimisation equation~\eqref{eq:min:biq:t:tree:radial}, we obtain
\begin{equation}
  4 \vec{e}_{i}^{T} \! \mat{\Lambda} \vec{e}_{i} + \vec{e}_{i}^{T} \! \mat{\beta} \vec{e}_{i} = 0
\end{equation}
or
\begin{equation}
  4 \lambda_{ii} + \beta_{ii} = 0,
\end{equation}
which  --- if we assume only scalar contributions ---  takes the form
\begin{equation}
  4 \lambda_{ii} = -\frac{1}{16 \pi^{2}} {\Bigg (} 72 \lambda_{ii}^{2} + 2 \sum_{j \neq i} \lambda_{ij}^{2} {\Bigg )}.
\label{eq:min:axis:scalars:only}
\end{equation}
The minimisation equation~\eqref{eq:min:axis:scalars:only} is a quadratic equation in $\lambda_{ii}$. Of the two solutions, one is dominated by $\lambda_{ij}$ and yields a perturbative value for $\lambda_{ii}$. We will see that it is a minimum of the effective potential.\footnote{The other solution, instead, is dominated by $\lambda_{ii}$ and results in a non-perturbative value of the coupling. We anticipate that such a solution corresponds to a maximum of the potential even if $\mathbb{B}<0$ as, in this case, the correction from $d\mathbb{B}/dt$ actually dominates.}

The tree-level mass matrix~\eqref{eq:m:sq:tree:biq} is given by
\begin{equation}
\begin{split}
  \mat{m}_{S}^{2} &= \diag(4 \mat{\Lambda} \vec{e}_{i}^{\circ 2}) v_{\varphi}^{2} 
  + 8 \mat{\Lambda} \circ (\vec{e}_{i} \vec{e}_{i}^{T}) \, v_{\varphi}^{2}
  \\
  &= \diag(4 \vec{\Lambda}_{i} 
  + 8 \lambda_{ii} \vec{e}_{i}) \, v_{\varphi}^{2},
\end{split}
\end{equation}
where $\vec{\Lambda}_{i}$ is the $i$th column of the coupling matrix $\mat{\Lambda}$. Since it is a diagonal matrix, the tree-level masses are simply $m_{i}^{2} = 12 \lambda_{ii} v_{\varphi}^{2} < 0$ and $m^{2}_{j \neq i} = 2 \lambda_{ij} v_{\varphi}^{2}$.
In order to calculate the quantum-corrected mass matrix, we use
\begin{align}
  2 \nabla_{\vec{\Phi}} \mathbb{B} &= 4 \vec{\Phi} \circ \mat{\beta} \vec{\Phi}^{\circ 2}
  = 4 v_{\varphi}^{3} \beta_{ii} \vec{e}_{i},
  \\ 
  \nabla_{\vec{\Phi}} t &= \frac{1}{v_{\varphi}} \vec{e}_{i}, 
  \\
   \nabla_{\vec{\Phi}} \nabla_{\vec{\Phi}}^{T} t &= 
  \frac{1}{v_{\varphi}^{2}} (\mat{I} -
   \vec{e}_{i} \vec{e}_{i}^{T}) + \frac{1}{2} \nabla_{\vec{\Phi}} \nabla_{\vec{\Phi}}^{T} \frac{\mathbb{A}}{\mathbb{B}},
\end{align}
and through eq.~\eqref{eq:mass:matrix:full:biq} find that the mass matrix is a diagonal matrix with entries
\begin{align}
  M_{i}^{2} &=  4 \beta_{ii} \, v_{\varphi}^{2}, 
  \\
  M_{j \neq i}^{2} &= 2 \lambda_{ij} v_{\varphi}^{2} + \beta_{ii} v_{\varphi}^{4} \partial_{\phi_{j}}^{2} t.
  \label{eq:axis:M:j}
\end{align}

\begin{figure}[tb]
\begin{center}
  \includegraphics{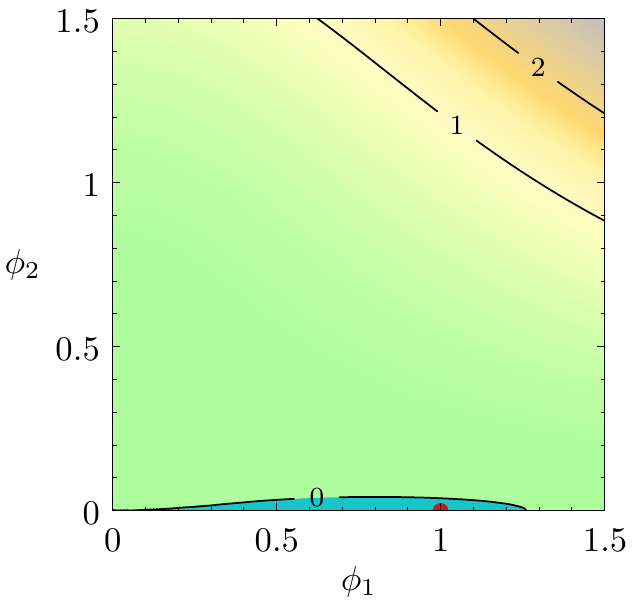}
  \caption{Contour plot of a potential with a minimum on an axis (red dot). The field values are in units of $v_{\varphi}$.}
\label{fig:axis:min:V}
\end{center}
\end{figure}

More concretely, suppose that the minimum of the potential~\eqref{eq:2:field:V} is on the $\phi_{1}$-axis. Then $\lambda_{1} < 0$ at the minimum and $\beta_{1} > 0$. The block form of the coupling matrix in eq.~\eqref{eq:block} is given by $\mat{\Lambda}_{11} = ( \lambda_{1} )$, $\mat{\Lambda}_{12} = ( \lambda_{12} )$, $\mat{\Lambda}_{22} = ( \lambda_{2} )$ and $\vec{\Phi}_{1} = ( v_{\varphi} )$. The copositivity of $\mat{\Lambda}_{22}$ thus requires $\lambda_{2} > 0$.
We have, neglecting $\lambda_{1}$ and $\lambda_{2}$, that
\begin{equation}
  M_{1}^{2} \approx \frac{\lambda_{12}^{2}}{2 \pi^{2}} v_{\varphi}^{2},
  \qquad
  M_{2}^{2} \approx 2 \lambda_{12} v_{\varphi}^{2} + M_{1}^{2}.
  \label{eq:min:axis:M:2}
\end{equation}
For perturbative couplings, the requirement $M_{2}^{2} > 0$ implies $\lambda_{12} > 0$.

An example of potential~\eqref{eq:2:field:V} with a minimum on the $\phi_{1}$-axis is shown in figure~\ref{fig:axis:min:V}, where we used $\lambda_{1} = -7.916 \times 10^{-3}$, $\lambda_{2} = 0.1$,  $\lambda_{12} = 0.5$ and the couplings have been evaluated at the radiative minimum for $\mu = \exp(-3/4) v_{\varphi}$. The field values are given in units of $v_{\varphi}$. The minimum (indicated by a red dot) lies in a basin of negative potential values. We have $M_{1} = 0.113 \, v_{\varphi}$ and $M_{2} = 1.006 \, v_{\varphi}$. The second term in \eqref{eq:min:axis:M:2} amounts to a $0.6 \%$ correction to $M_{2}$.

\subsection{Democratic minimum}
\label{subsec:min:dem}

The other case that can be analytically solved results from a ``democratic'' choice of the couplings: we set all self-couplings to a common value $\lambda_{ii} = \lambda_{s}$ and all portal couplings equal to each other $\lambda_{ij} = \lambda_{p}$, $i \neq j$. Hence,
\begin{equation}
  \mat{\Lambda} = \lambda_{s} \mat{I} + \frac{1}{2} \lambda_{p} (\vec{e} \vec{e}^{T} - \mat{I}),
\end{equation}
where $\vec{e} \vec{e}^{T} - \mat{I}$ is a matrix with zeros on the main diagonal and ones everywhere else. Provided that possible gauge and Yukawa couplings are also the same for all scalars, the symmetry of the problem determines the following solution:
\begin{equation}
  \vec{\Phi} = \frac{1}{\sqrt{n}} v_{\varphi} \vec{e}.
\label{eq:dem:sol}
\end{equation}
The normalisation factor $\frac{1}{\sqrt{n}}$ arises from $\vec{e}^{T} \vec{e} = n$, the number of scalar degrees of freedom.

It is easy to see that the solution~\eqref{eq:dem:sol} coincides with the lowest order solution to the minimisation equation~\eqref{eq:min:biq:t:tree:almost:flat:sol}: $\adj (\mat{\Lambda})$ has the same structure as $\mat{\Lambda}$ and $\adj(\mat{\Lambda}) \, \vec{e}$ is the vector of row sums of $\adj(\mat{\Lambda})$, which are all equal. This implies that in the democratic case $\nabla_{\vec{\Phi}^{\circ 2}} (\mathbb{A}/\mathbb{B}) = \vec{0}$, as can be analytically verified.

Inserting the solution~\eqref{eq:dem:sol} in the radial minimisation equation~\eqref{eq:min:biq:t:tree:radial}, we obtain
\begin{equation}
  4 \vec{e}^{T} \! \mat{\Lambda} \vec{e} + \vec{e}^{T} \! \mat{\beta} \vec{e} = 0.
\label{eq:symm:radial:min}
\end{equation}
Obviously, given the assumptions, the $\beta$-functions $\beta_{s}$ for the self-couplings are equal, and so are those for the portals, $\beta_{p}$. Eq.~\eqref{eq:symm:radial:min} thus takes the form
\begin{equation}
  4 \left( \lambda_{s} + \frac{n - 1}{2} \lambda_{p} \right)
  + \beta_{s} + \frac{n - 1}{2} \beta_{p} = 0.
\label{eq:symm:radial:min:comp}
\end{equation}

The tree-level mass matrix~\eqref{eq:m:sq:tree:biq} is given by
\begin{equation}
  \mat{m}^{2}_{S} = \frac{v_{\varphi}^{2}}{n} \left[ (12 \lambda_{s} + 2 (n - 3) \lambda_{p}) \mat{I} 
  + 4 \lambda_{p} \vec{e} \vec{e}^{T} \right],
\label{eq:mass:dem}
\end{equation}
where we used $\diag (\vec{e} ) = \mat{I}$.
The eigenvalues of the matrix~\eqref{eq:mass:dem} are given by the eigenvalues of the last term, shifted by the coefficient of the unit matrix. Since  $(\vec{e} \vec{e}^{T}) \vec{e} = n \vec{e}$, the eigenvalues of $\vec{e} \vec{e}^{T}$ are $n$, with multiplicity $1$, and $0$, with multiplicity $n - 1$.
That is, the tree-level masses along the minimum direction and in the orthogonal directions are given, respectively, by 
\begin{align}
  m_{\varphi}^{2} &= \frac{1}{n} [12 \lambda_{s} + 6 (n - 1) \lambda_{p}] v_{\varphi}^{2},
  \\ 
  m^{2}_{\perp} &= \frac{1}{n} [12 \lambda_{s} + 2 (n - 3) \lambda_{p}] v_{\varphi}^{2}.
\end{align}
 The minimum field vector~\eqref{eq:dem:sol} is an eigenvector of the tree-level mass matrix with $m_{\varphi}^{2} v_{\varphi}^{2} = \vec{\Phi}^{T} \mat{m}^{2}_{S} \vec{\Phi}$ and an eigenvector of the quantum-corrected mass matrix. Inserting the eigenvalue $m^{2}_{\perp}$ in the eigenvector equation of $\mat{m}^{2}_{S}$, we find that $\vec{e}^{T} \vec{n}_{\perp} = 0$, as required. The vector $\vec{n}_{\perp}$ has unit norm and is orthogonal to the VEV $\vec{\Phi}$ in field space.

The quantum-corrected masses at the minimum are 
\begin{equation}
  M_{\varphi}^{2} = -\frac{16}{n} \left( \lambda_{s} + \frac{n - 1}{2} \lambda_{p} \right) v_{\varphi}^{2}, 
  \qquad
  M_{\perp}^{2} = m^{2}_{\perp} + \frac{1}{4} M_{\varphi}^{2} v_{\varphi}^{2} \; \vec{n}_{\perp}^{T} (\nabla_{\vec{\Phi}} \nabla_{\vec{\Phi}}^{T} t) \vec{n}_{\perp},
\end{equation}
where we used $M_{\varphi}^{2} = 8 \mathbb{B}/v_{\varphi}^{2} = - 16 V/v_{\varphi}^{2}$ and the fact that both $\mat{\Lambda} \vec{\Phi}^{\circ 2}$ and $\mat{\beta} \vec{\Phi}^{\circ 2}$ are proportional to $\vec{e}$ and thus orthogonal to $\vec{n}_{\perp}$.

\begin{figure}[tb]
\begin{center}
  \includegraphics{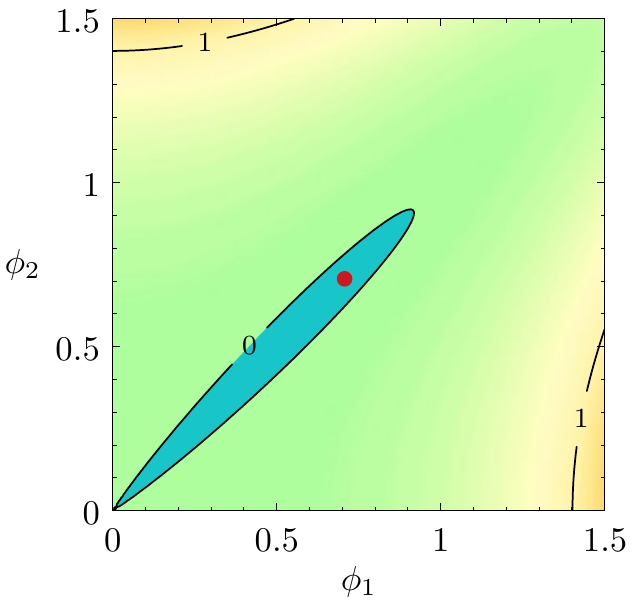}
  \caption{Contour plot of a potential with a democratic minimum (red dot). Field values are in units of $v_{\varphi}$.}
\label{fig:dem:min}
\end{center}
\end{figure}

For the two-field potential~\eqref{eq:2:field:V}, we have in the democratic case $\lambda_{1} = \lambda_{2}$ and the minimum solution is
\begin{equation}
  \vec{\Phi} = \frac{1}{\sqrt{2}} v_{\varphi} \begin{pmatrix} 1 \\ 1 \end{pmatrix}.
\end{equation}
An  example of democratic minimum for the potential~\eqref{eq:2:field:V} is shown in figure~\ref{fig:dem:min}, where we took $\lambda_{1} = \lambda_{2} = 0.244$ and $\lambda_{12} = -0.5$ in the minimum with $\mu = \exp(-3/4) \sqrt{2} v_{\varphi}$. The masses are $M_{\varphi} = 0.23 \, v_{\varphi}$ and $M_{\perp} = 1.43 \, v_{\varphi}$. The second term in $M_{\perp}$ amounts to a $1.5 \%$ correction.

\subsection{Iterative solution}
\label{subsec:example:iter}

To exemplify the iterative solution to the stationary point equation, we choose $\lambda_{1} = 0.06$, $\lambda_{2} = 1.125$ and $\lambda_{12} = -0.5$ at $\mu_{\rm in} = 1$ in the potential \eqref{eq:2:field:V}. The chosen values of the self-couplings differ by an order of magnitude to strongly depart from the democratic case discussed above.

First, we solve the RGEs to obtain the running couplings.  All the elements of $\adj (\mat{\Lambda})$ are positive in the neighbourhood of $\mu_{\rm in} = 1$, indicating that the potential may have a minimum lying on the field plane.\footnote{Of course, by running the couplings to small scales, the self-couplings could become negative (for instance $\lambda_{1}<0$ for $\mu < 10^{-10}$). However, this will not generate another minimum, as we shall see in subsection \ref{subsec:min:axis:plane}.}
We establish that the effective potential does have a minimum by studying the running of $\det (\mat{\Lambda})$: there is a flat direction at $\mu =  0.789$, below which the determinant is negative.

To begin the iteration procedure, we insert the lowest order solution \eqref{eq:min:biq:t:tree:almost:flat:sol} in the radial equation \eqref{eq:min:biq:t:tree:radial}, which we solve for $t_{0}$ and $\mat{\Lambda}(t_{0})$ simultaneously. Then we reparametrise our RGE solutions and solve $t(\varphi=v_{\varphi 0}, \mu=\mu_0)=0$ to determine $v_{\varphi 0}$. The first order iteration is obtained by inserting $\vec{\Phi}^{\circ 2}_{0} = v_{\varphi 0}^{2} \vec{n}^{\circ 2}_{0}$ into \eqref{eq:sol:1st:order}, solving the radial equation for $t_{1}$ (and $\mat{\Lambda}(t_{1})$), and finding $v_{\varphi 1}$ from $t_1$ after a further reparametrisation. The iteration procedure is repeated until sufficient precision is obtained.

The results after each iteration step are given in table~\ref{tab:i}, where we present the renormalisation scale $\mu$ at the minimum and the corresponding values of the couplings. In place of the field components $\phi_{1}$ and $\phi_{2}$, we give the radial coordinate $v_{\varphi}$ and the minimum vector angle $\cos \theta$. In addition, we display the masses and the mixing angle $\cos \alpha$, which is close, but not exactly equal, to $\cos \theta$. The first order correction is already within sight of the actual minimum and the second and third order give identical results to the given accuracy. The mass $M_{1}$ in the lowest order approximation is overestimated by $4.5 \%$, whereas the mass $M_{2}$ is $11 \%$ smaller than the true value at the minimum. Table~\ref{tab:i} also shows the values of the spectral radius $\rho(\mat{J})$ of the Jacobian given in the appendix \ref{sec:convergence}.  When it is less than unity at the fixed point, the iteration converges. 

\begin{figure}[tb]
\begin{center}
  \includegraphics{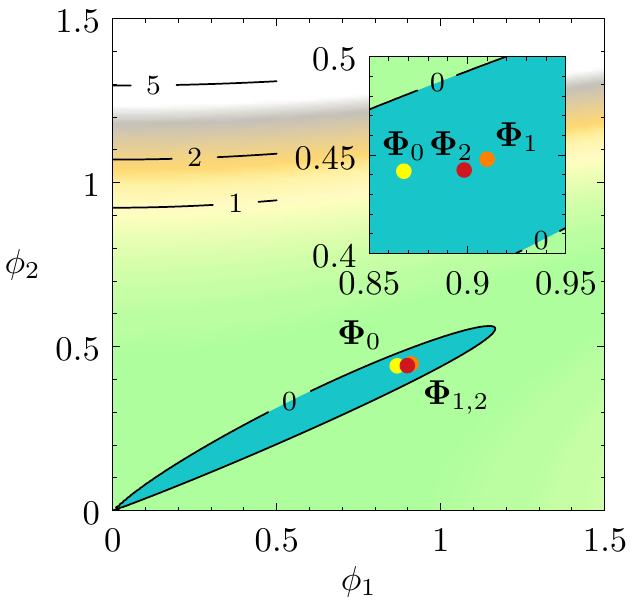}
  \caption{Iterative solution to the minimisation problem: contour plot of the potential. We show the 0th order (yellow dot), 1st order (orange) and 2nd order (red) iterations. Field values are in units of $v_{\varphi}$. The inset shows  the region near the minimum in detail.}
\label{fig:iter:min}
\end{center}
\end{figure}

\begin{table}[tb]
\centering
\begin{tabular}{|c|cccccccccc|}
\hline
Order & $\mu$ & $\lambda_{1}$  & $\lambda_{2}$  & $\lambda_{12}$ & $v_{\varphi}$ & $\cos \theta$ & $M_{1}$ & $M_{2}$ & $\cos \alpha$ & $\rho(\mat{J})$ \\
\hline 
$0$ & $0.6209$ & $0.0578$ & $0.903$ & $-0.474$ & $0.97$ & $0.8911$ & $0.198$ & $1.313$ & $0.9090$ & $1$  \\
$1$ & $0.6286$ & $0.0579$ & $0.907$ & $-0.474$ & $1.01$ & $0.8972$ & $0.192$ & $1.486$ & $0.9150$ & $0.93$  \\
$2$ & $0.6286$ & $0.0579$ & $0.907$ & $-0.474$ & $1.00$ & $0.8971$ & $0.190$ & $1.465$ & $0.9149$ & $0.93$ \\
\hline
\end{tabular}
\caption{\label{tab:i} Iterative solution to the minimum. The first iteration is already very close to the true minimum; to the given accuracy, the third iteration gives the same result as the second.}
\end{table}

Figure \ref{fig:iter:min} shows the iteration steps to the minimum overlaid on the contour plot of the potential $V$. The lowest order solution is indicated by a yellow dot, the first iteration by an orange dot and the second iteration by a red dot. The third iteration already coincides with the second one for all practical purposes.
The angle between the fixed point minimum vector and the lowest order solution \eqref{eq:min:biq:t:tree:almost:flat:sol} differs by less than a degree, but the length $v_{\varphi 0}$ of the lowest order solution differs from the actual length by $2.8\%$.

To analytically gauge the size of the corrections, we compare the two terms in the solution \eqref{eq:min:biq:t:tree:nonzero:final} for $\vec{\Phi}^{\circ 2}$ at first order,  neglecting the Goldstone mass for simplicity. The first term is
\begin{equation}
  \vec{\Phi}_{0}^{\circ 2} = \frac{V}{ \det(\mat{\Lambda})} \adj (\mat{\Lambda}) \vec{e} 
  = \frac{V}{ \det(\mat{\Lambda})} \frac{1}{v_{\varphi}^2} 
  \begin{pmatrix}
  2 \lambda_{1} - \lambda_{12}
  \\
  2 \lambda_{2} - \lambda_{12}
  \end{pmatrix}
\end{equation}
and the second term is
\begin{equation}
  \Delta \vec{\Phi}^{\circ 2} \equiv  
  \frac{V}{ \det(\mat{\Lambda})}\adj(\mat{\Lambda}) \nabla_{\vec{\Phi}^{\circ 2}} \frac{\mathbb{A}}{\mathbb{B}}
  = \frac{V}{ \det(\mat{\Lambda})} \frac{1}{v_{\varphi}^2} \frac{3 (\lambda_{1} - \lambda_{2}) \det(\mat{\Lambda})}{12 \lambda_{1} \lambda_{2} - 4 \lambda_{12} (\lambda_{1} + \lambda_{2}) + \lambda_{12}^{2}  
   }
   \begin{pmatrix}
   1
   \\
   -1
   \end{pmatrix}.
\end{equation}
Note that $\Delta \vec{\Phi}^{\circ 2}$ is only approximately equal to $\vec{\Phi}^{\circ 2}_{1} - \vec{\Phi}^{\circ 2}_{0}$, as the values of quartic couplings used in $\vec{\Phi}^{\circ 2}_{0}$ and $\vec{\Phi}^{\circ 2}_{1}$ are different. Because $\Delta \vec{\Phi}^{\circ 2}$ is orthogonal to $\vec{e}$, it is convenient to compare the projections $\vec{\Phi}^{\circ 2}_{0}$ and $\Delta \vec{\Phi}^{\circ 2}$ on $\vec{e}_{\perp} = (1,-1)^{T}$. Furthermore, we expect the iteration to converge when the ratio of norms
\begin{equation}
  \frac{\abs{\Delta \vec{\Phi}^{\circ 2}}}{\abs{\vec{\Phi}_{0}^{\circ 2}}}
  = \frac{6 \abs{\lambda_{1} - \lambda_{2}} \abs{\lambda_{12}^{2} - 4 \lambda_{1} \lambda_{2}}}{\abs{12 \lambda_{1} \lambda_{2} - 4 \lambda_{12} (\lambda_{1} + \lambda_{2}) + \lambda_{12}^{2}} \abs{2 \lambda_{1}^{2} - 2 \lambda_{1} \lambda_{12} + \lambda_{12}^{2} - 2 \lambda_{12} \lambda_{2} + 2 \lambda_{2}^{2}}}
\end{equation}
is less than unity. For the example at hand, $\vec{e}_{\perp}^{T} \Delta \vec{\Phi}^{\circ 2} / \vec{e}^{T}_{\perp} \vec{\Phi}_{0}^{\circ 2}$ is 3.4\% and $\abs{\Delta \vec{\Phi}^{\circ 2}}/\abs{\vec{\Phi}_{0}^{\circ 2}}$ is 1.7\% at the minimum. Note that as $\abs{\Delta \vec{\Phi}^{\circ 2}} \propto \abs{\lambda_{1} - \lambda_{2}}$, the correction is zero when the self-couplings are equal, which is exactly the case of the democratic minimum (section \ref{subsec:min:dem}).

Now that we have found the minimum, let us test our method for the inverse problem of section \ref{subsec:inv:problem} and find the coupling matrix $\mat{\Lambda}$, starting with the minimum vector $\vec{\Phi}$ and the mass matrix $\mat{M}^{2}_{S}$. The lowest order approximation \eqref{eq:inv:Lambda:0} to the coupling matrix gives $\lambda_{1} = 0.0579$, $\lambda_{12} = -0.4904$ and $\lambda_{2} = 1.152$, so  the relative error with the true coupling matrix is $\norm{\mat{\Lambda} - \mat{\Lambda}_{0}} / \norm{\mat{\Lambda}} = 0.25$. The renormalisation scale in the minimum is obtained from solving the equation $t(\varphi = v_{\varphi} = \sqrt{\vec{\Phi}^{T}\vec{\Phi}}, \mu) = 0$ for $\mu = 0.728 v_{\varphi}$. Beginning with these values, we calculate $\mathbb{B}$ and $t$ and use eq. \eqref{eq:inv:Lambda} to obtain $\mat{\Lambda}$ via iteration. It takes $10$ iterations to bring the relative error $\norm{\mat{\Lambda}_{i+1} - \mat{\Lambda}_{i}} / \norm{\mat{\Lambda}_{i+1}}$ below $1\%$, after which the relative error with the true coupling matrix is $\norm{\mat{\Lambda} - \mat{\Lambda}_{10}} / \norm{\mat{\Lambda}}$ is about $3\%$ (further iteration does not improve the precision).

\section{Multiple minima}
\label{sec:multi:min}

\begin{figure}[t]
\begin{center}
  \includegraphics{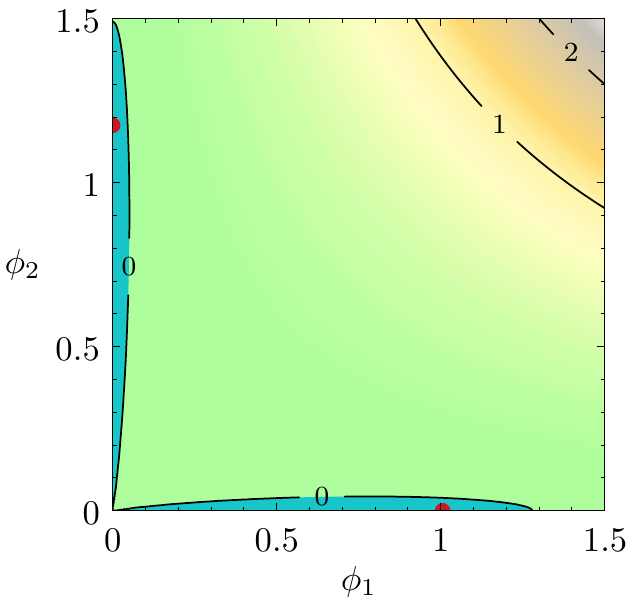}
\caption{An effective potential with two minima, each on a field space axis. Field values are in units of the $\phi_{1}$-axis minimum $v_{\varphi 1}$.}
\label{fig:axes:mins}
\end{center}
\end{figure}

The lion's share of literature considers only the presence of one minimum in the effective potential. Although it has been pointed out that more minima could co-exist \cite{Chataignier:2018kay}, to our best knowledge, a detailed discussion of the implications of such a possibility is still absent.

The simplest way to construct an effective potential with several separate minima is to arrange them in different same-dimensional subspaces of the field space, e.g. on field space axes (along the lines of the discussion in section~\ref{subsec:min:subspace}). For example, we discussed a minimum on a field space axis in section \ref{subsec:min:axis} within the two-field model with the potential~\eqref{eq:2:field:V}. In  this scenario, in order to have a minimum on \emph{each} axis, it is clear that \emph{both} self-couplings $\lambda_{1}$ and $\lambda_{2}$ must be negative below certain scales, resulting in separate minima that need not be degenerate ($v_{\varphi 1}\neq v_{\varphi 2}$). Such a situation is depicted in figure~\ref{fig:axes:mins} for the values $\lambda_{1} = -7.94 \times 10^{-4}$, $\lambda_{2}= -1.29 \times 10^{-3}$  and $\lambda_{12} = 0.5$ at $\mu = \exp (-3/4) v_{\varphi 1}$, where $v_{\varphi 1}$ is the VEV of the $\phi_{1}$ field. In this case, it is the positive portal coupling $\lambda_{12}$ that separately drives both the self-couplings through zero. The value of the potential in the $\phi_{1}$ minimum is $V = -7.92 \times 10^{-4} v_{\varphi 1}^{4}$ and the particle masses are $M_{1} = 0.113 v_{\varphi 1}$ and $M_{2} = 1.006 v_{\varphi 1}$. In the $\phi_{2}$ minimum with $v_{\varphi 2} = 1.165 \, v_{\varphi 1}$ (at $\mu = 0.550 v_{\varphi 1}$), the value of the potential is $V = -8.04 \times 10^{-4}  v_{\varphi 2}^{4} = -1.46 \times 10^{-3} v_{\varphi 1}^{4}$ and the particle masses are $M_{1} = 0.132 v_{\varphi 2} = 0.153 v_{\varphi 1}$ and $M_{2}  = 0.870 v_{\varphi 2} = 1.01 v_{\varphi 1}$. In a similar manner, it is possible to arrange two or more minima situated on different field planes and so on. This way for constructing effective potentials with multiple minima is robust and can effortlessly be implemented in models with more than two fields.

In more involved cases, several minima lie in the \emph{same} subspace of field space (an axis or a plane, for instance). In order to have two minima on the same field space axis, the self-coupling must increase, decrease and increase again. Its $\beta$-function must therefore first become negative and then positive again, crossing zero twice. Similarly, if two or more minima are on a field plane or in a larger subspace, $\mathbb{B}$ has to run through zero between every pair of stationary points, as shown in section \ref{subsec:mass:matrix}.

\subsection{Toy model}
\label{subsec:toy}

To demonstrate these more complicated examples, we use a simple toy model with two complex scalar fields, $\Phi_{1}$ and $\Phi_{2}$, and two Weyl fermions $\psi_{0}$ and $\psi_{1}$. We take $\Phi_{2}$ to be complex merely in order not introduce extra numerical coefficients in the potential. The fermions transform as $\psi_{0} \sim (0,0)$ and $\psi_{1} \sim (1,0)$ under the $\mathbb{Z}_{2}^{2}$ symmetry of the potential.\footnote{See appendix \ref{sec:fermion:mass:biquadratic} for how the indices correspond to representations. In the scheme of appendix \ref{sec:fermion:mass:biquadratic}, scalars have a different enumeration from fermions by which the two scalars should have been indexed $0$ and $1$. For ease of comparison with the simple two-field model, however, we use as indices their representation numbers $2^{0} = 1$ and $2^{1} = 2$.} The $\Phi_{1}$ and $\psi_{0}$ fields transform non-trivially under a non-Abelian gauge group, for concreteness as $SU(2)$ doublets, while $\Phi_{2}$ and $\psi_{1}$ transform as singlets. The Lagrangian is then
\begin{equation}
   \mathcal{L} = -\frac{1}{4} \tr (F_{\mu\nu} F^{\mu\nu}) + i \bar{\psi}_{0} \slashed{D} \psi_{0}
 + i \bar{\psi}_{1} \slashed{\partial} \psi_{1} + D_{\mu} \Phi_{1}^{\dagger} D^{\mu} \Phi_{1} + \partial_{\mu} \Phi_{2}^{\dagger} \partial^{\mu} \Phi_{2} - \mathcal{L}_{Y} - V,
\end{equation}
with the Yukawa interactions and the scalar potential given by
\begin{align}
  \mathcal{L}_{Y} &= y \bar{\psi}_{0} \Phi_{1} \psi_{1} + \text{H.c.},
  \\
  V &= \lambda_{1} \abs{\Phi_{1}}^{4} + \lambda_{12} \abs{\Phi_{1}}^{2} \abs{\Phi_{2}}^{2} 
  + \lambda_{2} \abs{\Phi_{2}}^{4}.
\end{align}
We parametrise
\begin{equation}
  \Phi_{1} = 
  \begin{pmatrix}
   \frac{\phi_{1} + i \phi_{2}}{\sqrt{2}}
   \\
    \frac{\phi_{3} + i \phi_{4}}{\sqrt{2}}
  \end{pmatrix},
  \qquad
  \Phi_{2} =
  \frac{\phi_{5} + i \phi_{6}}{\sqrt{2}},
\end{equation}
and by defining $\frac{1}{2} \varphi_{1}^{2} \equiv \abs{\Phi_{1}}^{2} = \frac{1}{2} (\phi_{1}^{2} + \phi_{2}^{2} + \phi_{3}^{2} + \phi_{4}^{2})$ and $\frac{1}{2} \varphi_{2}^{2} \equiv \abs{\Phi_{2}}^{2} = \frac{1}{2} (\phi_{5}^{2} + \phi_{6}^{2})$, we can write the potential as
\begin{equation}
  V =  \frac{1}{4} \lambda_{1} \varphi_{1}^{4} + \frac{1}{4} \lambda_{12} \varphi_{1}^{2} \varphi_{2}^{2} 
  + \frac{1}{4} \lambda_{2} \varphi_{2}^{4},
\end{equation}
so the coupling matrix of the field vector $\vec{\Phi} = (\varphi_{1}, \varphi_{2})^{T}$ is 
\begin{equation}
  \mat{\Lambda} = 
  \frac{1}{4}
  \begin{pmatrix} 
    \lambda_{1} & \frac{1}{2} \lambda_{12} \\
    \frac{1}{2} \lambda_{12} & \lambda_{2}
  \end{pmatrix}.
\end{equation}
The tree-level field-dependent scalar masses are given by
\begin{align}
  m_{1,2}^{2} &= \frac{1}{4} \left[ (6 \lambda_{1} + \lambda_{12}) \varphi_{1}^{2} + (6 \lambda_{2} + \lambda_{12}) \varphi_{2}^{2} \right] 
  \notag
  \\
  &\pm \frac{1}{4} \sqrt{(\lambda_{12} - 6 \lambda_{1})^{2} \varphi_{1}^{4} +2 [7 \lambda_{12}^{2} + 6 \lambda_{12} (\lambda_{1} + \lambda_{2}) - 36 \lambda_{1} \lambda_{2}] \varphi_{1}^{2} \varphi_{2}^{2} + (\lambda_{12} - 6 \lambda_{2})^{2} \varphi_{2}^{4}},
  \\
  m_{3,4,5}^{2} &= \lambda_{1} \varphi_{1}^{2} + \frac{1}{2} \lambda_{12} \varphi_{2}^{2},
  \\
  m_{6}^{2} &= \lambda_{2} \varphi_{2}^{2} + \frac{1}{2} \lambda_{12} \varphi_{1}^{2},
\end{align}
where $m_{1,2}$ are the masses of the two mass eigenstates arising from the singlet-doublet mixing, $m_{3,4,5}$ are $SU(2)$ Goldstone masses and $m_{6}$ is the mass of $\phi_{6}$, the Goldstone of a global $U(1)$ symmetry.
There is one Dirac fermion $\Psi = (\psi_0, \psi_1)$ mass eigenstate with mass
\begin{equation}
  m_{\Psi}^{2} = \frac{1}{2} y^{2} \varphi_{1}^{2}
\end{equation}
and one massless Weyl fermion.
The field-dependent mass of the $SU(2)$ vector bosons arises from the covariant derivative (appendix \ref{sec:gauge:biquadratic}) and is
\begin{equation}
  m_{V}^{2} = \frac{1}{4} g^{2} \varphi_{1}^{2}.
\end{equation}

The one-loop $\beta$-functions and anomalous dimensions were calculated with \textsc{sarah} \cite{Staub:2013tta} and are given by
\begin{align}
  16 \pi^{2} \beta_{g^{2}} &= -\frac{41}{3} g^{4},
  \\
  16 \pi^{2} \beta_{y^{2}} &= \frac{1}{2} y^{2} (10 y^{2} - 9 g^{2}),
  \\
  16 \pi^{2} \beta_{\lambda_{1}} &= \frac{9}{8} g^{4} - 9\lambda_{1} g^{2} + 24 \lambda_{1}^{2} + \lambda_{12}^{2} + 4 \lambda_{1} y^{2} - 2 y^{4},
  \\
  16 \pi^{2} \beta_{\lambda_{2}} &= 20 \lambda_{2}^{2} + 2 \lambda_{12}^{2},
  \\
  16 \pi^{2} \beta_{\lambda_{12}} &= \lambda_{12} \left( -\frac{9}{2} g^{2} + 2 y^{2} + 12 \lambda_{1} + 4 \lambda_{12} + 8 \lambda_{2} \right),
  \\
  16 \pi^{2} \gamma_{\Phi_{1}} &= - \frac{9}{4} g^{2} + y^{2},
  \\
  16 \pi^{2} \gamma_{\Phi_{2}} &= 0.
\end{align}
The quantities that enter the $t$ parameter are
\begin{equation}
  64 \pi^{2} \mathbb{B} =  \sum_{i=1}^{6} m^{4}_{i} + 9 m^{4}_{V} - 4 m^{4}_{\Psi}
\end{equation}
and
\begin{equation}
  64 \pi^{2} V^{(1)} = \sum_{i=1}^{6} m^{4}_{i} \left(\ln \frac{\abs{m^{2}_{i}}}{\mu^{2}} - \frac{3}{2} \right)
  + 9 m^{4}_{V} \left(\ln \frac{m^{2}_{V}}{\mu^{2}} - \frac{5}{6} \right) - 4 m^{4}_{\Psi} \left(\ln \frac{m^{2}_{\Psi}}{\mu^{2}} - \frac{3}{2} \right),
\end{equation}
where we take absolute values of the scalar $m^{2}_{i}$ to deal with the spurious negative Goldstone masses.

\subsection{Two minima on an axis}
\label{subsec:two:min:axis}

We start by discussing how the toy model effective potential can admit two minima on the same axis of the scalar field space.

In order to generate two minima on the $\phi_{1}$-axis, the running self-coupling $\lambda_{1}$ must be negative in two different intervals. Let us set $\lambda_{1}$ negative at low scales ($t = 0$). Its $\beta$-function $\beta_{\lambda_{1}}$ must then be positive at low scales, negative at intermediate scales, and positive at high scales to ensure vacuum stability. In short, it must run through zero twice. For negligible $\lambda_{1}$ contributions, we have that
\begin{equation}
  16 \pi^{2} \beta_{\lambda_{1}} \approx \frac{9}{8} g^{4} + \lambda_{12}^{2} - 2 y^{4}.
\end{equation}
The positive contributions from scalars and gauge bosons must dominate at low scales and high scales, while at intermediate scales, instead, the negative Yukawa contributions must prevail. In order to keep the minimum on the axis we need a positive portal coupling $\lambda_{12}$. As the portal coupling grows with the running, it will dominate the large scale region. The non-Abelian gauge coupling $g$, on the contrary, keeps $\beta_{\lambda_{1}}$ positive at low scales but decreases with the scale. An almost constant Yukawa coupling then ensures that  $\beta_{\lambda_{1}}$ is negative at intermediate scales.

\begin{figure}[tb]
\begin{center}
\includegraphics{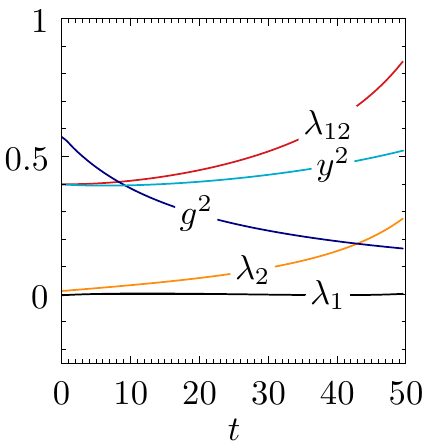}~~
\includegraphics{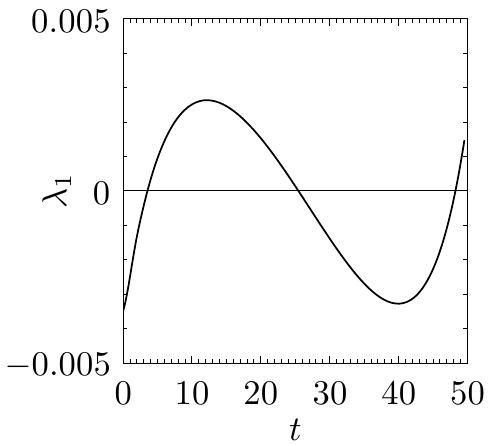}~~
\includegraphics{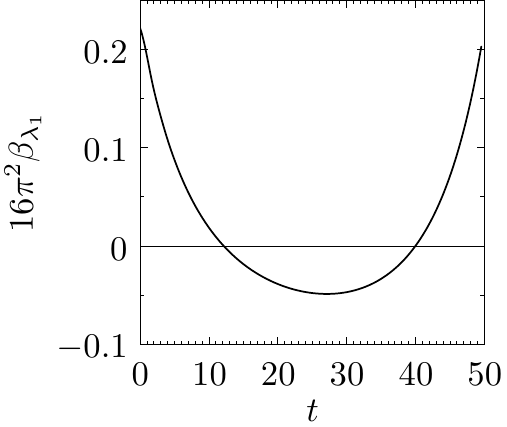}
\caption{Running of couplings that generates two minima on a field space coordinate axis. Left panel: Running of the scalar self-couplings $\lambda_{1}$ and $\lambda_{2}$, the portal coupling $\lambda_{12}$, the gauge coupling $g$ and the Yukawa coupling $y$. Middle panel: The self-coupling $\lambda_{1}$ is negative in two regions and the effective potential has a minimum in each. Right panel: Such running of $\lambda_{1}$ is due to a $\beta_{\lambda_{1}}$ that is positive at low and high scales and negative at intermediate scales.}
\label{fig:axis:mins:running}
\end{center}
\end{figure}

Since we already know the minimum direction, we find the scales of both minima from the radial eq. \eqref{eq:min:biq:t:tree:radial}. We choose the values of the couplings in the first minimum with unit VEV $v_{\varphi \text{I}}$ (at the renormalisation scale $\mu = 0.264 v_{\varphi \text{I}}$) as $\lambda_{1} = -1.962 \times 10^{-4}$, $\lambda_{2} = 0.01832$, $\lambda_{12} = 0.4003$, $g = 0.7039$ and $y = 0.6287$. The running of couplings is shown as a function of $t = \ln  (\mu / 0.01 v_{\varphi \text{I}})$ in the left panel of figure~\ref{fig:axis:mins:running}. The middle panel shows the running of the self-coupling $\lambda_{1}$ and the right panel the running of its $\beta$-function $\beta_{\lambda_{1}}$.

The value of the potential in the first minimum is $V = -4.906 \times 10^{-5} v_{\varphi \text{I}}^{4}$. The mass matrix is diagonal. The mass of the pseudo-Goldstone of the scale invariance is $M_{1} = 0.0280 v_{\varphi \text{I}}$. Since the global $U(1)$ associated by the singlet $\Phi_{2}$ is not broken, both its components have the same mass $M_{2} = 0.447 v_{\varphi \text{I}}$. The field is not rescaled in the first minimum, because there we choose our initial conditions,  but in the second minimum we have to take into account anomalous dimensions. We find that the renormalisation scale $\mu = 7.098 \times 10^{18} v_{\varphi \text{I}}$ in the second minimum, so $\phi_{1}$ is scaled by $\exp( \Gamma_{1} (t)) = 1.0541$, while $\phi_{2}$ does not scale.
The VEV in the second minimum is therefore $v_{\varphi \text{II}} = \exp(-\Gamma_{1} (t)) 1.282 \times 10^{19} v_{\varphi \text{I}} = 1.216 \times 10^{19} v_{\varphi \text{I}}$, the value of the potential is $V = -7.170 \times 10^{-5}  v_{\varphi \text{II}}^{4} = -1.582 \times 10^{72} v_{\varphi \text{I}}^{4}$ and the particle masses are $M_{1} = 0.0322 v_{\varphi \text{II}} = 3.925 \times 10^{17} v_{\varphi \text{I}}$ and $M_{2} = 0.6611 v_{\varphi \text{II}} = 8.057 \times 10^{18} v_{\varphi \text{I}}$. As expected, a saddle point separates the two minima.

\subsection{Two minima in a plane}
\label{subsec:two:min:plane}

Another scenario we can explore with our toy model has two minima in the field plane, i.e. two minima, where both scalars have non-zero VEVs. We can use a similar configuration of couplings as in the previous example. The exception is that now the portal coupling $\lambda_{12}$ must be negative. However, a negative $\lambda_{12}$ tends to run slower, so we need a larger $\lambda_{2}$ to make its absolute value large at high scales. For that reason, the second minimum is generated right before the couplings hit a Landau pole. Whereas this issue should be addressed in realistic models, it is irrelevant for the purpose of our example.

\begin{figure}[tb]
\begin{center}
\hspace{-2mm}
\includegraphics{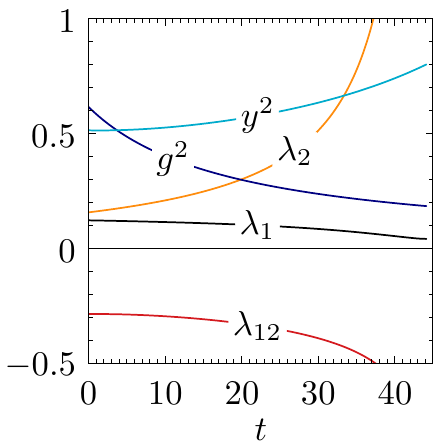}~~
\includegraphics{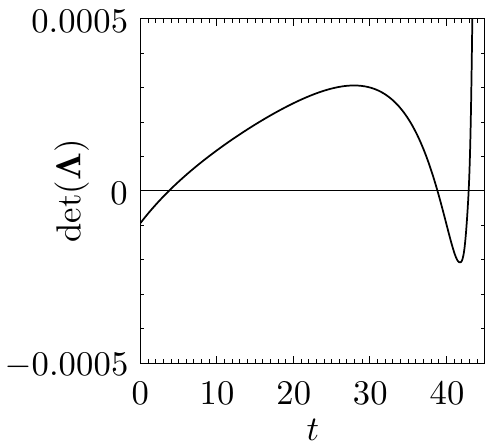}~~
\includegraphics{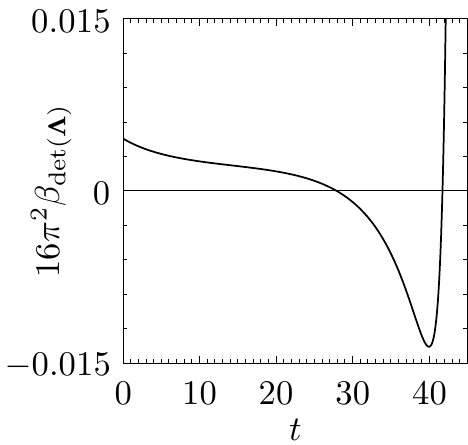}
\caption{Running of couplings that generates two minima on the field plane. Left panel: Running of the scalar self-couplings $\lambda_{1}$ and $\lambda_{2}$, the portal coupling $\lambda_{12}$, the gauge coupling $g$ and the Yukawa coupling $y$. Middle panel: The determinant of the coupling matrix is negative in two regions and the effective potential has a minimum in each. Right panel: The required running of $\det(\mat{\Lambda})$ is arranged with a $\beta_{\det(\mat{\Lambda})}$ that is positive at low and high scales but negative at intermediate scales.}
\label{fig:plane:mins:running}
\end{center}
\end{figure}

In order to forbid minima on the axes, we require $\lambda_{1} > 0$ and $\lambda_{2} > 0$. To check whether two minima appear, we track the running of $\det(\mat{\Lambda})$. This quantity must run thrice through zero, similarly to the $\lambda_{1}$ in the case of two minima on the $\phi_{1}$-axis treated in the previous subsection. We also need $\adj(\mat{\Lambda}) > 0$, which holds on the whole range considered, since $\lambda_{1} > 0$, $\lambda_{2} > 0$ and $-\lambda_{12} > 0$ for all $t$. Figure \ref{fig:plane:mins:running} shows the running of couplings, of $\det(\mat{\Lambda})$ and of $\beta_{\det (\mat{\Lambda})}$ as a function of  $t = \ln  (\mu / 0.01 v_{\varphi \text{I}})$.

We find the first minimum via the iteration procedure in section \ref{subsec:sol:it}. We give everything in units of its radial VEV $v_{\varphi \text{I}}$. The values of couplings in the first minimum are $\lambda_{1} = 0.1195$, $\lambda_{2} = 0.1718$, $\lambda_{12} = -0.2872$, $g = 0.7208$ and $y = 0.7169$ at $\mu = 0.231 v_{\varphi \text{I}}$. The first minimum is at $\phi_{1} = 0.738 v_{\varphi \text{I}}$, $\phi_{2} = 0.674 v_{\varphi \text{I}}$. The value of the potential is $V = -3.873 \times 10^{-5} v_{\varphi \text{I}}^{4}$. The masses of the singlet-doublet mixed mass eigenstates are $M_{1} = 0.0248 v_{\varphi \text{I}}$ and $M_{2} = 0.530 v_{\varphi \text{I}}$. The imaginary part of $\Phi_{2}$ is the Goldstone of the breaking of a global $U(1)$.

The iteration does not converge for the second minimum due to large quantum corrections in the vicinity of the Landau pole. Therefore, we use the full expression for $t_*$ to improve the effective potential and find the minimum numerically. The second minimum is at $\mu = 3.27 \times 10^{16} v_{\varphi \text{I}}$, which scales the $\phi_{1}$ component by $\exp(\Gamma_{1}(t)) = 1.0267$, while $\phi_{2}$ does not scale. The second minimum is then at $\phi_{1} = 0.919 v_{\varphi \text{II}}$, $\phi_{2} = 0.333 v_{\varphi \text{II}}$ with $v_{\varphi \text{II}} = 3.17 \times 10^{16} v_{\varphi \text{I}}$. The value of the potential at the second minimum is  $V = -5.943 \times 10^{-5} v_{\varphi \text{II}}^{4} = -6.001 \times 10^{61} v_{\varphi \text{I}}^{4}$ and the particle masses are $M_{1} = 0.809 v_{\varphi \text{II}} = 2.56 \times 10^{16} v_{\varphi \text{I}}$ and $M_{2} = 0.0620 v_{\varphi \text{II}} = 1.96 \times 10^{15} v_{\varphi \text{I}}$. Again, a saddle point separates the two minima.

\subsection{One minimum on an axis, one on the plane}
\label{subsec:min:axis:plane}

Is it also possible to arrange one minimum on an axis and another on the field plane? With only two scalars it is impossible, though such a configuration may be achievable considering three or more fields. With two scalars, in fact, the running portal coupling $\lambda_{12}$ keeps its sign. To have a minimum on the plane, we need a negative portal coupling, but this is in contradiction with the $\lambda_{12} > 0$ requirement imposed to keep the orthogonal field mass \eqref{eq:min:axis:M:2} positive for the minimum on an axis. Even in the case of a nearly vanishing portal, other couplings cannot produce a positive mass via quantum corrections (as in the second term of eq.~\eqref{eq:axis:M:j}) without non-perturbative values.

\section{Conclusions}
\label{sec:concl}

We have considered the minimisation of effective potentials with multiple scalar fields by way of a novel matrix formalism in which calculations are presented in a compact and intuitive manner. Our approach is easy to implement in any modern computer algebra system.  The matrix formalism could also be useful for studying the symmetry breaking due to negative mass-squared terms.

To demonstrate the formalism in a clear way, we consider an RG-improving method such that the one-loop corrections vanish due to a specific field-dependent choice of the renormalisation scale \cite{Chataignier:2018aud}. The resulting effective potential has the tree-level form with running couplings evaluated at that scale. In the main, we are concerned with popular biquadratic potentials, but many of our results apply to more general potentials via tensor algebra. The approach is not limited to the particular improvement procedure adopted and can be adapted to other choices of the renormalisation scale, such as $\mu = \varphi$,  through the straightforward changes detailed in appendix~\ref{sec:impro}. Dimensionful terms arising from thermal corrections or from a non-minimal coupling to gravity in a curved background can also be treated in our matrix formalism because they respect the biquadratic symmetry of the scalar potential. In the latter case, for example, the renormalisation scale can depend on the curved background \cite{Markkanen:2018bfx}.\footnote{However, unless the scalars are spectator fields, it is necessary to provide an additional relation linking the evolution of the background to the scalar field values due to the back-reaction of the field evolution on the background and on the same effective mass term sourced by the non-minimal coupling. The matter is further complicated by the choice of frame (Jordan or Einstein) to describe the dynamics, as well as by the details pertaining to the computation of the RG-improved potential in the two frames, which could possibly lead to inequivalent results.}

Our first main result concerns the stationary point equation~\eqref{eq:min:biq:t:tree:nonzero:final} for the scalar potential, derived within a matrix formulation. At the lowest order \eqref{eq:min:biq:t:tree:almost:flat:sol}, the minimum solution has the same functional form as the corresponding flat direction \cite{Kannike:2019upf}, but differs from it due to quantum corrections. Neglecting those, the lowest order solution reproduces the Gildener-Weinberg approximation. In general, the minimum equation can only be solved iteratively. We show that the iteration converges and discuss in detail the first order correction with a semi-analytical treatment that goes beyond purely numerical considerations. The lowest order solution is exact in two special cases, which we solve analytically: a minimum on a field space axis and a ``democratic'' minimum, where all scalar fields have equal self-couplings and identical portal couplings.

We stress that the formalism is not limited to Gildener-Weinberg minima. In fact, given a generic effective potential with a radiatively generated minimum, since at large field values we must have $V > 0$ and in the minimum necessarily it is $V < 0$, there is always a point in field space with $V = 0$ at some intermediate value of the radial coordinate. For Gildener-Weinberg solutions, this flat direction and the minimum are aligned. Differently, when the tree-level and one-loop contributions are comparable (as typical of Coleman-Weinberg solutions), the ``flat direction'' and the minimum may not be aligned at all. As our criterion for the existence of a minimum given in \eqref{eq:min:conds} depends only on the quartic couplings, our formalism can address both mentioned cases.

Once the direction in field space of the effective potential minimum is determined, one can calculate the quantum-corrected mass matrix \eqref{eq:mass:matrix:full:biq}. Its positive-definiteness requires the potential and the quartic coupling matrix to satisfy a set of conditions in the minimum \eqref{eq:min:conds}. We also solve the inverse problem, obtaining expressions for the quartic couplings and the $\beta$-functions at the minimum given the desired minimum field vector of VEVs, physical masses and mixing angles for the fields.

The second main result concerns the exploration of effective potentials with several physical minima. In the literature, only a single Gildener-Weinberg type of minimum is generally considered, although more minima can be present. For example, a two-scalar biquadratic effective potential can admit two minima located on different axes of the field space (figure \ref{fig:axes:mins}), two minima on the same axis (figure \ref{fig:axis:mins:running}), or two minima in a common field plane at different distances from the origin (figure \ref{fig:plane:mins:running}). We provide semi-analytical criteria to determine whether an effective potential has more than one minimum. In such cases, the running quartic couplings must satisfy the same set of conditions \eqref{eq:min:conds} at two different scales or in orthogonal field subspaces. Examples are given in a minimal toy model with two complex scalar fields, two Weyl fermions and an $SU(2)$ gauge symmetry meant to resemble the SM.

Because most of the models in the literature are concerned with biquadratic potentials, we expect that our methods will facilitate the exploration of such effective potentials. Considering classically scale-invariant effective potentials with more than one minimum may open new directions in phenomenological studies of phase transitions and cosmic gravitational wave signals.


\appendix

\section{\texorpdfstring{$\beta$}{Beta}-function matrix}
\label{sec:beta:matrix}

We derive the $\beta$-function for the quartic coupling matrix $\mat{\Lambda}$ of the biquadratic potential~\eqref{eq:pot:biq}. In the absence of gauge or Yukawa interactions, the anomalous dimensions at one-loop level are zero, and eq.~\eqref{eq:C-Z} implies that
\begin{equation}
    \sum_{i} \sum_{j} \beta_{ij} \frac{\partial V^{(0)}}{\partial \lambda_{ij}} = \sum_{i} \sum_{j} \beta_{ij} \phi_{i}^{2} \phi_{j}^{2} 
  = (\vec{\Phi}^{\circ 2})^{T} \! \mat{\beta} \vec{\Phi}^{\circ 2} = \frac{1}{32 \pi^{2}} \Str \mat{m}^{4},
\end{equation}
with the scalar contribution given by
\begin{equation}
\begin{split}
  \Str \mat{m}_S^{4} 
  &= \tr \left( 16 [\diag ( \mat{\Lambda} \vec{\Phi}^{\circ 2} )]^{2} 
  + 32  \diag ( \mat{\Lambda} \vec{\Phi}^{\circ 2} ) \; [\mat{\Lambda} \circ (\vec{\Phi} \vec{\Phi}^{T})]
  \right.
  \\
  &\left. + 32 [\mat{\Lambda} \circ (\vec{\Phi} \vec{\Phi}^{T})] \; \diag ( \mat{\Lambda} \vec{\Phi}^{\circ 2} ) 
  + 64 [\mat{\Lambda} \circ (\vec{\Phi} \vec{\Phi}^{T})]^{2} \right).
\end{split}
\end{equation}
Calculating the traces and bringing $\Str \mat{m}_S^{4}$ into the same form as $(\vec{\Phi}^{\circ 2})^T \!\mat{\beta}  \vec{\Phi}^{\circ 2}$, we obtain
\begin{equation}
\begin{split}
  16 \pi^{2} \mat{\beta} &= 32 \mat{\Lambda}^{\circ 2} + 8 \mat{\Lambda}^{2}
  + 16 \mat{\Lambda} \Diag (\mat{\Lambda}) + 16 \Diag (\mat{\Lambda}) \, \mat{\Lambda}
  \\
  &+ \text{gauge and Yukawa contributions},
\end{split}
\label{eq:d:Lambda:dt}
\end{equation}
where $\Diag(\mat{\Lambda}) \equiv \mat{\Lambda} \circ \mat{I}$ is the diagonal matrix with the diagonal of $\mat{\Lambda}$. The $\beta$-function \eqref{eq:d:Lambda:dt} matches the general results \cite{Machacek:1983tz,Machacek:1983fi,Machacek:1984zw,Luo:2002ti} restricted to biquadratic potentials.

The $\beta$-function for the determinant can be computed, using Jacobi's formula, as
\begin{equation}
\begin{split}
  16 \pi^{2} \frac{d \det (\mat{\Lambda})}{dt} &= 16 \pi^{2} \tr \adj (\mat{\Lambda}) \mat{\beta} 
  \\
  &= \tr \left[ 32 \adj (\mat{\Lambda}) \mat{\Lambda}^{\circ 2} + 40 \det(\mat{\Lambda}) \mat{\Lambda} \right] \\
  &+ \text{gauge and Yukawa contributions.}
  \label{eq:beta:det}
\end{split}
\end{equation}

We can find the $\beta$-function for the adjugate by differentiating the defining relation
\begin{equation}
  \mat{\Lambda} \adj(\mat{\Lambda}) = \det (\mat{\Lambda}) \mat{I},
\end{equation}
yielding
\begin{equation}
  \det (\mat{\Lambda}) \frac{d \adj(\mat{\Lambda}) }{dt} 
  = \adj(\mat{\Lambda}) \frac{d \det(\mat{\Lambda}) }{dt} - \adj(\mat{\Lambda}) \frac{d\mat{\Lambda}}{dt} \adj(\mat{\Lambda}).
\label{eq:d:adj:dt}
\end{equation}

\section{Biquadratic symmetry-preserving Yukawa couplings}
\label{sec:fermion:mass:biquadratic}

The Yukawa Lagrangian for left-handed complex Weyl fermions $\psi_{i}$ is given by
\begin{equation}
  -\mathcal{L}_{Y} = \frac{1}{2} Y^{a}_{jk} \phi_{a} \psi_{j} \epsilon \psi_{k}
  + \frac{1}{2} Y^{a *}_{jk} \phi_{a} \psi_{j}^{\dagger} \epsilon \psi_{k}^{\dagger},
\label{eq:Yukawa}
\end{equation}
where $\epsilon = i \sigma_{2}$ is the fermion metric. In general, Yukawa couplings do not respect the $\mathbb{Z}_{2}^{n}$ symmetry of a biquadratic potential.

In order for a Yukawa coupling involving the field $\phi_{a}$ to respect the symmetry, the fermion bilinear $\psi_{j} \epsilon \psi_{k}$ must be odd under $(\mathbb{Z}_{2})_{a}$. Hence, one of the fermions must be odd under $(\mathbb{Z}_{2})_{a}$ and \emph{both} fermions may also be odd under a subset of the other $\mathbb{Z}_{2}$ symmetries.

An even parity corresponds to the $\mathbb{Z}_{2}$ charge $0$, since $+1 = e^{i 0 \pi}$; similarly an odd parity corresponds to the $\mathbb{Z}_{2}$ charge $1$, since $-1 = e^{i 1 \pi}$. It is most convenient to start counting the Yukawa matrix indices from zero and number a scalar or fermion by treating its representation as a binary number\footnote{We note that a mixed radix number system could be used to number representations of more complicated discrete symmetries whose factors are not only $\mathbb{Z}_{2}$, but also higher $\mathbb{Z}_{N}$ symmetries.} --- with the order of bits reversed, so that the numbering of $(\mathbb{Z}_{2})_{a}$ begins from the left: $0 = (0, \ldots, 0)$, $1 = (1, 0 \ldots, 0)$, $2 = (0, 1, 0, \ldots, 0)$, $3 = (1, 1, 0, \ldots, 0)$, \ldots, $2^{n} - 1 = (1, \ldots, 1)$. There can be several fermions in the same representation --- for instance if the fermion carries colour --- whose Yukawa couplings we sum together for simplicity. The $n$ scalars are naturally numbered from $2^{0}$ to $2^{n-1}$, where only one bit equals $1$ and others are $0$, so the scalar $\phi_{a}$ has number $2^{a}$. A Yukawa coupling between $\phi_{a}$, $\psi_{i}$ and $\psi_{j}$ can thus be non-zero only if $2^{a} \oplus i \oplus j = 0$, where $\oplus$ is the bitwise \textsc{xor}. That this yields an even parity is easily seen from $0 \oplus 0 \oplus x = 1 \oplus 1 \oplus x = x$ for $x = 0~\text{or}~1$. A Yukawa coupling of $\phi_{a}$ with fermion $\psi_{i}$ and fermion $\psi_{j}$ is therefore given by
\begin{equation}
  Y^{a}_{ij} = y^{a}_{ij} \delta_{(2^{a} \oplus i \oplus j) 0}.
\label{eq:Y:biq}
\end{equation}
We can see that the diagonal $i = j$ of the Yukawa matrix is zero.

For example, if $n = 2$, then the possible $\mathbb{Z}_{2}^{2}$ representations are $0 = (0,0)$, $1 = (1,0)$, $2 = (0,1)$ and $3 = (1,1)$. Then $\phi_{0}$ in representation $2^{0} = 1 = (1,0)$ can have Yukawa couplings with pairs of fermions in representations $0 = (0,0)$ and $1 = (1,0)$, or $2 = (0,1)$ and $3 = (1,1)$. At the same time, $\phi_{1}$ in representation $2^{1} = 2 = (0,1)$ can have Yukawa couplings with pairs of fermions in representations $0 = (0,0)$ and $2 = (0,1)$, or $1 = (1,0)$ and $3 = (1,1)$.

The field-dependent fermion mass matrix is given by
\begin{equation}
  (\mat{m}_{F})_{ij} = \sum_{a} Y^{a}_{ij} \phi_{a},
\end{equation}
so
\begin{equation}
  (\mat{m}_{F}^{2})_{ij} = \sum_{a,b,k} Y^{\dagger a}_{ik} Y^{b}_{kj} \phi_{a} \phi_{b}.
\end{equation}
Then we have, for example,
\begin{equation}
\begin{split}
   \tr \mat{m}_{F}^{4} &= \sum_{a,b,c,d,i,j,k,l}  Y^{\dagger a}_{ij} Y^{b}_{jk}
   Y^{\dagger c}_{kl} Y^{d}_{li} \phi_{a} \phi_{b} \phi_{c} \phi_{d}
   \\
   &= \sum_{a,b,c,d,i,j,k,l} y^{a *}_{ji} y^{b}_{jk} y^{c *}_{lk} y^{d}_{li}
   \delta_{(2^{a} \oplus i \oplus j)0} \delta_{(2^{b} \oplus j \oplus k)0}
   \delta_{(2^{c} \oplus k \oplus l)0} \delta_{(2^{d} \oplus l \oplus i)0} 
   \phi_{a} \phi_{b} \phi_{c} \phi_{d},
\end{split}
\label{eq:tr:mF:4}
\end{equation}
which implies that the indices of non-zero terms in the sum~\eqref{eq:tr:mF:4} are given by the solutions of
\begin{equation}
  0 = 2^{a} \oplus i \oplus j = 2^{b} \oplus j \oplus k = 2^{c} \oplus k \oplus l = 2^{d} \oplus l \oplus i.
\label{eq:xor:tr:mF:4:indices:eqs}
\end{equation}
The bitwise \textsc{xor} is commutative and associative and satisfies
\begin{equation}
  u \oplus v = w \iff v \oplus w = u \iff u \oplus w = v.
\end{equation}
Eqs.~\eqref{eq:xor:tr:mF:4:indices:eqs} therefore imply
\begin{align}
  2^{a} \oplus i &= 2^{b} \oplus k,
\label{eq:xor:tr:mF:4:indices:eqs:2:1}
  \\
  2^{d} \oplus i &= 2^{c} \oplus k,
\label{eq:xor:tr:mF:4:indices:eqs:2:2}
\end{align}
and, in turn, that $2^{a} \oplus 2^{b} = 2^{c} \oplus 2^{d}$, $2^{a} \oplus 2^{d} = 2^{b} \oplus 2^{c}$ and $2^{a} \oplus 2^{c} = 2^{b} \oplus 2^{d}$, that is $a \oplus b = c \oplus d$, $a \oplus d = b \oplus c$ and $a \oplus c = b \oplus d$. We see that assuming e.g. $a = b$ implies $c = d$, and so on. On the other hand, if we set e.g. $a = b = c \neq d$, the equations~\eqref{eq:xor:tr:mF:4:indices:eqs} have no solution.

Let us show that there are no non-biquadratic terms in $\tr \mat{m}_{F}^{4}$. Having a non-biquadratic term requires that at least two $\phi_{i}$ differ, e.g. $a \neq b$ without loss of generality. Then if either $a$ or $b$ equals $c$ or $d$, we saw that the term must be biquadratic. Therefore all $a$, $b$, $c$ and $d$ must be different from each other to have a chance to produce a non-biquadratic term. For that, we need $n \geq 4$.

We can assume for the sake of concreteness, again without loss of generality, that
\begin{equation}
  a = 0, \quad b = 1, \quad c = 2, \quad d = 3.
\end{equation}
The least significant bit of $2^{a}$ is then $(2^{a})_{1} = 1$, while $(2^{b})_{1} = (2^{c})_{1} = (2^{d})_{1} = 0$. From eqs.~\eqref{eq:xor:tr:mF:4:indices:eqs} we have for the least significant bits of fermion indices that
\begin{equation}
  i_{1} \oplus j_{1} = 1, \quad k_{1} \oplus j_{1} = 0, \quad k_{1} \oplus l_{1} = 0,
  \quad i_{1} \oplus l_{1} = 0.
\label{eq:first:bits}
\end{equation}
Assuming a concrete value for $j_{1}$, e.g. $j_{1} = 0$, the values of all the other first bits are determined and eq.~\eqref{eq:first:bits} yields a contradiction. Therefore $\tr \mat{m}_{F}^{4}$ is a biquadratic function of the scalars as expected. The proof trivially generalises to $n > 4$.

\section{Gauge contributions to the \texorpdfstring{$\beta$}{beta}-function}
\label{sec:gauge:biquadratic}

The scalar fields are coupled to gauge bosons via their covariant derivatives, given by
\begin{equation}
  D_{\mu} \vec{\Phi} = \partial_{\mu} \vec{\Phi} - i g V_{\mu}^{A} \mat{\theta}^{A} \vec{\Phi},
\end{equation}
where the representations $\mat{\theta}^{A}$ of the group generators are Hermitian matrices. Because we have decomposed complex scalars in terms of their real components, the $\mat{\theta}^{A}$ matrices are purely imaginary and antisymmetric. The field-dependent vector boson mass matrix is 
\begin{equation}
  (\mat{m}_{V}^{2})_{AB} = g^{2} \vec{\Phi}^{T} \mat{\theta}^{A} \mat{\theta}^{B} \vec{\Phi}
  =  \frac{1}{2} g^{2} \vec{\Phi}^{T} \{ \mat{\theta}^{A}, \mat{\theta}^{B} \} \vec{\Phi}
\end{equation}
and 
\begin{equation}
  \tr \mat{m}_{V}^{4} = \frac{1}{4} g^{4} \sum_{A,B} \vec{\Phi}^{T} \{ \mat{\theta}^{A}, \mat{\theta}^{B} \} \vec{\Phi} \; \vec{\Phi}^{T} \{ \mat{\theta}^{A}, \mat{\theta}^{B} \} \vec{\Phi}.
\end{equation}

\section{Anomalous dimensions}
\label{sec:anom:dim}

At one-loop level in the Landau gauge, the anomalous dimension of scalars are given by
\begin{equation}
  16 \pi^{2} \gamma_{ab} = 2 \kappa Y_{2}^{ab}(S) - 3 g^{2} C_{2}^{ab}(S),
\end{equation}
where $\kappa$ is $1/2$ for Weyl and $1$ for Dirac fermions \cite{Luo:2002ti} and
\begin{align}
  Y_{2}^{ab}(S) &= \frac{1}{2} \tr (\mat{Y}^{\dagger a} \mat{Y}^{b} + \mat{Y}^{\dagger b} \mat{Y}^{a}),
  \\
  C_{2}^{ab}(S) &= \sum_{A} \mat{\theta}^{A} \mat{\theta}^{A} = C_{2}(S)^{a} \delta_{ab},
\label{eq:anom:dim:gauge:indices}
\end{align}
where $C_{2}(S)^{a}$ is the quadratic Casimir operator  $C_{2}(S)$. In the presence of several gauge groups $g^{2} C_{2}(S) \to \sum_{k} g_{k}^{2} C_{2}^{k} (S)$.

We see from eq. \eqref{eq:anom:dim:gauge:indices} that the gauge contribution is always diagonal, while in general the Yukawa contribution is not. We will now specialise to Yukawa couplings that obey the $\mathbb{Z}_{2}^{n}$ symmetry of the potential, as detailed in appendix~\ref{sec:fermion:mass:biquadratic}. We have, for Yukawa couplings of Weyl fermions given by eq. \eqref{eq:Yukawa}, that
\begin{equation}
\begin{split}
  Y_{2}^{ab}(S) &= \frac{1}{2} \tr (\mat{Y}^{\dagger a} \mat{Y}^{b} + \mat{Y}^{\dagger b} \mat{Y}^{a})
  \\
  &= \frac{1}{2} \sum_{i,j} (y^{a *}_{ij} y^{b}_{ij} + y^{b *}_{ij} y^{a}_{ij}) \delta_{(2^{a} \oplus i \oplus j) 0}  \delta_{(2^{b} \oplus i \oplus j) 0}
  \\
  &= \delta_{ab} \sum_{i,j} \abs{y^{a}_{ij}}^{2} \delta_{(2^{a} \oplus i \oplus j) 0},
\end{split}
\end{equation}
where we used $\delta_{(2^{a} \oplus i \oplus j) 0}  \delta_{(2^{b} \oplus i \oplus j) 0} = \delta_{ab} \delta_{(2^{a} \oplus i \oplus j) 0}$.

Hence, the matrix of anomalous dimensions for a model with $\mathbb{Z}_{2}^{n}$ symmetry is diagonal and given by
\begin{equation}
  16 \pi^{2} \gamma_{ab} = \Bigg[ \sum_{i,j} \abs{y^{a}_{ij}}^{2} \delta_{(2^{a} \oplus i \oplus j) 0} - 3 g^{2} C_{2}(S) \Bigg] \delta_{ab}.
\end{equation}

\section{Convergence of the iterative solution}
\label{sec:convergence}

A criterion for the convergence of the iteration is that the spectral radius of the Jacobian of the RHS of eq.~\eqref{eq:min:biq:t:tree:nonzero:final} at the fixed point be less than unity. The spectral radius of a matrix is given by the largest absolute value of its eigenvalues. The Jacobian is 
\begin{align}
  \mat{J} &= \nabla_{\vec{\Phi}^{\circ 2}} \left[ 2 \frac{V}{\det (\mat{\Lambda})} \adj (\mat{\Lambda}) \nabla_{\vec{\Phi}^{\circ 2}} t \right]^{T}
  \notag
  \\
  &= 2 \frac{\nabla_{\vec{\Phi}^{\circ 2}} V}{\det (\mat{\Lambda})} [ \adj (\mat{\Lambda}) 
  \nabla_{\vec{\Phi}^{\circ 2}} t ]^{T}
 - 2 \frac{V}{\det (\mat{\Lambda})^{2}} \frac{d \det (\mat{\Lambda})}{dt} 
  \nabla_{\vec{\Phi}^{\circ 2}} t \; [ \adj (\mat{\Lambda}) \nabla_{\vec{\Phi}^{\circ 2}} t ]^{T}
  \notag
  \\
  &+ 2 \frac{V}{\det (\mat{\Lambda})} \nabla_{\vec{\Phi}^{\circ 2}} t \; 
  \left[ \frac{d \adj (\mat{\Lambda})}{dt} \nabla_{\vec{\Phi}^{\circ 2}} t \right]^{T}
  + 2 \frac{V}{\det (\mat{\Lambda})} 
  \left( \nabla_{\vec{\Phi}^{\circ 2}} \nabla_{\vec{\Phi}^{\circ 2}}^{T} t \right) 
  \adj (\mat{\Lambda})
  \notag
  \\
  &= 2 \frac{V}{\det (\mat{\Lambda})} \left[ \nabla_{\vec{\Phi}^{\circ 2}} t \nabla_{\vec{\Phi}^{\circ 2}}^{T} t \left( -\frac{1}{\det (\mat{\Lambda})} 
  \frac{d \det (\mat{\Lambda})}{dt}  \adj (\mat{\Lambda})
  + \frac{d \adj (\mat{\Lambda})}{dt} \right) 
 + \left( \nabla_{\vec{\Phi}^{\circ 2}} \nabla_{\vec{\Phi}^{\circ 2}}^{T} t \right) 
  \adj (\mat{\Lambda}) 
  \right]
  \notag
  \\
  &= 2 \frac{V}{\det (\mat{\Lambda})} \left[ -\frac{1}{\det (\mat{\Lambda})} 
  \nabla_{\vec{\Phi}^{\circ 2}} t \, \nabla_{\vec{\Phi}^{\circ 2}}^{T} t \adj (\mat{\Lambda}) \mat{\beta} + \nabla_{\vec{\Phi}^{\circ 2}} \nabla_{\vec{\Phi}^{\circ 2}}^{T} t
   \right] \adj (\mat{\Lambda}),
\label{eq:iter:jacob}
\end{align}
where we have used the minimisation equation $\vec{0} = \nabla_{\vec{\Phi}^{\circ 2}} V$ and eq.~\eqref{eq:d:adj:dt}. In addition, the radial minimum condition~\eqref{eq:min:biq:t:tree:radial} yields
\begin{equation}
  \det (\mat{\Lambda}) = -\frac{1}{4} \frac{\nabla_{\vec{\Phi}^{\circ 2}}^{T} t 
  \adj(\mat{\Lambda}) \mat{\beta} \adj(\mat{\Lambda}) \nabla_{\vec{\Phi}^{\circ 2}} t}{\nabla_{\vec{\Phi}^{\circ 2}}^{T} t \adj(\mat{\Lambda}) \nabla_{\vec{\Phi}^{\circ 2}} t},
\end{equation}
and we have that
\begin{align}
  \nabla_{\vec{\Phi}^{\circ 2}} t &= \frac{1}{2} \frac{1}{\vec{e}^{T} \vec{\Phi}^{\circ 2}} \vec{e} + \frac{1}{2} \nabla_{\vec{\Phi}^{\circ 2}} \frac{\mathbb{A}_{0}}{\mathbb{B}_{0}},
  \\
  \nabla_{\vec{\Phi}^{\circ 2}} \nabla_{\vec{\Phi}^{\circ 2}}^{T} t &= -\frac{1}{2} 
  \frac{1}{(\vec{e}^{T} \vec{\Phi}^{\circ 2})^{2}} \vec{e} \vec{e}^{T}
  + \frac{1}{2} \nabla_{\vec{\Phi}^{\circ 2}} \nabla_{\vec{\Phi}^{\circ 2}}^{T} \frac{\mathbb{A}_{0}}{\mathbb{B}_{0}}.
\end{align}
Thus at zeroth order, for example, we have
\begin{equation}
  \mat{J} = \frac{1}{\vec{e}^{T} \adj (\mat{\Lambda}) \vec{e}} \vec{e} \vec{e}^{T}
  \left[ 2 \frac{\vec{e}^{T} \! \adj (\mat{\Lambda}) \vec{e}}{\vec{e}^{T} \! \adj (\mat{\Lambda}) \mat{\beta} \adj (\mat{\Lambda}) \vec{e}} \adj (\mat{\Lambda}) \mat{\beta} - \mat{I} \right] \adj{\mat{\Lambda}}.
\end{equation}

\section{Formalism for other choices of improvement}
\label{sec:impro}

We briefly repeat the derivation of the minimisation equation, its solution and the mass matrix in section~\ref{sec:V:eff:min} for a generic value of the renormalisation scale $\mu = \mathcal{M}$. Common choices are captured in $\mathcal{M}^2 = \vec{e}_{\mathcal{M}}^T \vec{\Phi}^{\circ 2}$, where $\vec{e}_{\mathcal{M}}$ is a constant vector. In particular, $\mu = \varphi$ is given by $\vec{e}_{\mathcal{M}} = \vec{e}$.

The one-loop potential can be written as
\begin{equation}
\begin{split}
    V^{(1)} &= 
 \mathbb{A} + \mathbb{B} \ln \frac{\mathcal{M}^{2}}{\mu^{2}},
\end{split}
\label{eq:V:(1):gen}
\end{equation}
where 
\begin{align}
  \mathbb{A}  &= \frac{1}{64 \pi^{2}} \Str \mat{m}^{4} \left(\ln \frac{\mat{m}^{2}}{\mathcal{M}^{2}} - \mat{C} \right), 
\label{eq:A:gen}
\\
 \mathbb{B} &= \frac{1}{64 \pi^{2}} \Str \mat{m}^{4}.
\label{eq:B:gen}
\end{align}

By setting $\mu = \mathcal{M}$, the effective potential becomes
\begin{equation}
  V(t) = V^{(0)}(t) + \mathbb{A},
  \label{eq:Veff}
\end{equation}
where $V^{(0)}(t)$ is the tree-level potential with field-dependent effective couplings, and the running parameter is given by
\begin{equation}
  t = \frac{1}{2} \ln \frac{\mathcal{M}^{2}}{\mathcal{M}_{0}^{2}}.
\end{equation}

The stationary point equation is
\begin{equation}
  \vec{0} = \nabla_{\vec{\Phi}} V = 4 \vec{\Phi} \circ \mat{\Lambda} \vec{\Phi}^{\circ 2} 
  + \nabla_{\vec{\Phi}} \mathbb{A} + 2 \mathbb{B} \nabla_{\vec{\Phi}} t.
  \label{eq:minmf}
\end{equation}
The radial stationary point equation is obtained by projecting Eq.~\eqref{eq:minmf} along the field vector and has the same form for any choice of $\mathcal{M}$:
\begin{equation}
  0  =  \vec{\Phi}^{T} \nabla_{\vec{\Phi}} V = 4 V^{(0)}(t) + 4 \mathbb{A} + 2 \mathbb{B}
= 4 V + 2 \mathbb{B}.
\label{eq:min:rad}
\end{equation}
Using  the chain rule \eqref{eq:chain}, we can  write Eq.~\eqref{eq:minmf} as
\begin{equation}
  \vec{0} = \vec{\Phi} \circ \left( 4 \mat{\Lambda} \vec{\Phi}^{\circ 2} + \frac{1}{\mathcal{M}^{2}} 2 \mathbb{B} \, \vec{e}_{\mathcal{M}} + 2 \nabla_{\vec{\Phi}^{\circ 2}} \mathbb{A} \right).
  \label{eq:min:vector:2}
\end{equation}

We solve eq.~\eqref{eq:min:vector:2} for $\vec{\Phi}\neq \vec{0}$. An implicit equation for the solution is given by 
\begin{equation}
   \vec{\Phi}^{\circ 2} = \frac{\adj{(\mat{\Lambda})}}{\det (\mat{\Lambda})} 
   \left(\frac{V}{\mathcal{M}^{2}} \vec{e}_{\mathcal{M}} - \frac{1}{2} \nabla_{\vec{\Phi}^{\circ 2}} \mathbb{A} \right),
   \label{eq:sol:impl:gen}
\end{equation}
where we used the radial eq.~\eqref{eq:min:rad}.

 To lowest order, we neglect the last term in eq.~\eqref{eq:sol:impl:gen} and take $V \approx V^{(0)}$, yielding
\begin{equation}
\begin{split}
 \vec{\Phi}^{\circ 2} &= \frac{1}{\det( \mat{\Lambda} )} \frac{V}{\mathcal{M}^{2}} \adj( \mat{\Lambda} ) 
 \vec{e}_{\mathcal{M}}
 \\
 &=  \frac{\mathcal{M}^{2}}{\vec{e}_{\mathcal{M}}^{T} \! \adj (\mat{\Lambda}) \vec{e}_{\mathcal{M}}} \adj (\mat{\Lambda}) \vec{e}_{\mathcal{M}},
\end{split}
  \label{eq:sol:expl}
\end{equation}
where we used eq.~\eqref{eq:V:lowest} with $\vec{e} \to \vec{e}_{\mathcal{M}}$.
For the first iteration, we insert the lowest-order solution \eqref{eq:sol:expl} in \eqref{eq:sol:impl:gen}.

The mass matrix around the minimum is then given by 
\begin{equation}
  \mat{M}_{S}^{2} = \nabla_{\vec{\Phi}} \nabla_{\vec{\Phi}}^{T} V 
 = \mat{m}^{2}_{S} + 2 \nabla_{\vec{\Phi}} \mathbb{B} \nabla_{\vec{\Phi}}^{T} t
+  2 \nabla_{\vec{\Phi}} t \nabla_{\vec{\Phi}}^{T} \mathbb{B}
 + 2 \mathbb{B} \nabla_{\vec{\Phi}} \nabla_{\vec{\Phi}}^{T} t
 + \nabla_{\vec{\Phi}} \nabla_{\vec{\Phi}}^{T} \mathbb{A},
\end{equation}
where $\mat{m}^{2}_{S}$ is the tree-level scalar mass matrix. Notice that the term proportional to $\mathbb{B}$ is canceled by a similar term resulting from  $\nabla_{\vec{\Phi}} \nabla_{\vec{\Phi}}^{T} \mathbb{A}$.
For a minimum in a field subspace, like in section~\ref{subsec:min:subspace}, the mass matrix still obtains a block-diagonal form due to the discrete symmetry of the potential.

\acknowledgments

We are grateful to Bogumi\l{}a \'Swie\.{z}ewska, Antonio Racioppi, Marco Piva, Carlo Marzo, Alexandros Karam and Damiano Anselmi for useful discussions. This work was supported by the Estonian Research Council grants PRG356 and PRG434, and by the European Regional Development Fund and programme Mobilitas Pluss grant MOBTT5, and the ERDF CoE program project TK133.


\bibliography{Veff_min}

\end{document}